\documentclass[fleqn,usenatbib]{mnras}

\usepackage[T1]{fontenc}
\usepackage{ae,aecompl}

\usepackage{graphicx}	
\usepackage{amsmath}	
\usepackage{amssymb}	
\usepackage{courier}
\usepackage{xcolor}
\usepackage{arydshln}
\usepackage{upgreek}
\usepackage{blindtext}
\usepackage{lipsum}
\newcommand\error{\varepsilon}
\newcommand{\sigmat}{\tilde{\sigma}}
\newcommand{\dif}{{\rm d}}
\newcommand{\rhot}{\tilde{\rho}}
\newcommand{\DPsieff}{\Delta\Psi_{\rm eff}}
\newcommand{\Psieff}{\Psi_{\rm eff}}
\newcommand{\DPsiz}{\Delta\Psi_z}

\title[SN Disequilibrium in Model Residual]{Residuals of an Equilibrium Model for the Galaxy Reveal a State of Disequilibrium in the Solar Neighborhood}

\author[H. Li et al.]{
Haochuan Li\thanks{E-mail: haochuan.li@queensu.ca}
and Lawrence M. Widrow\thanks{E-mail: widrow@queensu.ca}
\\
Department of Physics, Engineering Physics and Astronomy, Queen’s University, Kingston K7L 3X5, Canada}

\date{Accepted XXX; Received YYY; In original form ZZZ}

\pubyear{2021}

\begin{document}
\label{firstpage}
\pagerange{\pageref{firstpage}--\pageref{lastpage}}
\maketitle

\begin{abstract}
We simultaneously model the gravitational potential and phase space distribution function (DF) of giant stars near the Sun using the {\it Gaia} DR2 radial velocity catalog. We assume that the Galaxy is in equilibrium and is symmetric about both the spin axis of the disk and the Galactic midplane. The potential is taken as a sum of terms that nominally represent contributions from the gas disk, stellar disk, bulge, and dark matter halo. Our DF model for the giants comprise two components to account for a mix of thin and thick disk stars. The DF for each component is described by an analytic function of the energy, the spin angular momentum, and the vertical energy, in accord with Jeans theorem. We present model predictions for the radial and vertical forces within $\sim 2\,{\rm kpc}$ of the Sun, highlighting the rotation curve and vertical force profile in the Solar Neighbourhood. Finally, we show residuals for star counts in the $R-z$ and $z-v_z$ planes as well as maps of the mean radial and azimuthal velocities in the $z-v_z$ plane. Using our model for the potential, we also examine the star count residuals in action-frequency-angle coordinates. The {\it Gaia} phase spirals, velocity arches, some of the known moving groups and bending modes appear as well-defined features in these maps.
\end{abstract}

\begin{keywords}
Galaxy:kinematics and dynamics - Galaxy: Solar Neighborhood - Galaxy: disc - Galaxy: structure - Galaxy:evolution
\end{keywords}


\section{Introduction}\label{intro}

Our understanding of the mass distribution in galaxies invariably comes from models of the gravitational potential, which in turn are informed by observations of kinematic tracers and assumptions about their orbits. However, constraints on the mass distribution from tracers tend to be highly non-local, especially for external galaxies. For example, a galaxy's rotation curve is governed by the cumulative, roughly spherically-averaged mass distribution. By contrast, \citet{Oort1932} showed that for the Milky Way one can probe the local gravitation potential and hence local mass distribution by considering the vertical motions of stars in the Solar Neighbourhood. The framework he laid out has been the basis for numerous analyses in the intervening years and is known as the Oort Problem.

In general, stars in Milky Way's disk follow orbits that are determined by the mean Galactic gravitational field rather than close encounters with other stars. Their dynamical state is therefore described by a single particle distribution function (DF), which obeys the collisionless Boltzmann equation (CBE) and Poisson's equation \citep{GalacticDynamics}.  In the Oort Problem, one typically assumes that the Galaxy is in dynamical equilibrium and that both the DF and the potential are symmetric about the spin axis and Galactic midplane. The analysis then proceeds by one of several methods based on low-order moments of the CBE, namely the continuity and Jeans equations, or the Jeans theorem, which states that a DF for an equilibrium system can be written in terms of isolating integrals of motion.

Modeling the local potential and DF is technically challenging even when a high degree of symmetry is imposed. Approaches based on the Jeans theorem require a third integral of motion since models for the DF based on the two readily available integrals, namely the energy and the spin angular momentum, are inadequate to describe the physics of disk dynamics \citep{GalacticDynamics}. On the other hand, approaches based on moments of the CBE invariably include only the the mass continuity equation (zeroth moment) and Jeans equations (first moment). They therefore require additional assumptions to close the system of moment equations. A common strategy, which dates back to Oort's original work, is to assume that the local stellar dynamics in the vertical and in-plane directions decouple. Over the years, various methods have been devised to exploit this strategy and estimate the systematic errors that arise from the breakdown of its core assumption \citep{KG, holmberg2000, Garbari2012, Zhang2013, read2014, Xia2016, Guo2020}.

In \citet{PaperI} (hereafter Paper I) we simultaneously modelled the stellar DF and gravitational potential using kinematic measurements from the radial velocity catalog of {\it Gaia}'s Second Data Release (hereafter GDR2). We included only stars within $\sim1.5\,{\rm kpc}$ of the midplane whose Galactocentric radii were within $500\,{\rm pc}$ of the Solar Circle. The analysis in Paper I was strictly one-dimensional: we assumed that the DF for the tracers was a function of the vertical energy and that the potential was a function of $z$. 
A novel feature of Paper I was that it modelled the full $z-v_z$ DF and potential simultaneously and therefore one could diagnose departures from equilibrium from the model residuals. Over the past decade, numerous examples of disequilibrium in vertical structure of the Solar Neighborhood have been uncovered using data from various surveys such as RAVE \citep{RAVE}, SDSS \citep{SDSS}, LAMOST \citep{lamost} and \textit{Gaia} \citep{gaiadr2_summary}. For instance, the vertical number count profile showed a distinct pattern of North-South asymmetries of order $\sim10\%$ and with length scales on the order of hundreds of parsecs \citep{widrow2012,yanny2013, bennett2019, salomon2020}. Similarly, the bulk vertical velocity and velocity dispersion showed evidence for bending and breathing motions of the disk in the direction perpendicular to the midplane \citep{widrow2012, carlin2013, williams2013, carrillo2018}. Beyond the Solar Circle, the disk also appeared to be corrugated and warped \citep{ Xu2015, Sch2018, friske2019d, Poggio2018, Poggio2020}.

Perhaps the most intriguing manifestations of disequilibrium in the Solar Neighbourhood are the phase spirals discovered by \citet{antoja2018} using data from GDR2. These spirals appear in maps of number counts, mean radial velocity and mean vertical velocity as projected onto the vertical phase space, i.e. the $z-v_z$ plane. While the origin of the phase spirals is still a matter of debate, a promising idea is that they are partially phase-mixed perturbations caused by a disturbance in the disk, which itself may have been due by a passing satellite (see e.g. \citealt{Gomez2013, antoja2018, Binney2018, Darling2019, blandhawthorn2019, Laporte2018, laporte2019, Li2020, bennett2022}). In Paper I the number count phase spiral appeared as a residual of the DF. Since we also modelled the potential, we were able to transform the residual map to $\Omega_z-\theta_z$ coordinates where $\Omega_z$ is the vertical frequency and $\theta_z$ is the angle variable associated with the vertical action. In these coordinates, the spiral transformed to parallel, diagonal bands whose slope provided an estimate for the age of the perturbation that created it.
 
The most significant shortcoming of Paper I was its strict adherence to the 1D approximation. To illustrate this problem, we divided our sample into hot and cold sub-populations as defined by the in-plane kinetic energy. If the 1D approximation was strictly valid, the two sub-populations would have led to the same inferred potential and force. In fact, the inferred vertical force from these two sub-populations differed by $10\%-30\%$. In this work, we go beyond the 1D approximation by considering models for the full 6D DF and 3D potential while still retaining the assumptions of equilibrium, axisymmetry, and mirror symmetry about the midplane. We model the potential as the sum of contributions from a spherical dark matter halo, a point mass bulge, a razor-thin gas disk and a stellar disk. 
Our model for the DF of the tracer population includes two components to account for contributions from thin and thick disks. Following \citet{Kuijken_Dubinski_1995}, we model each component as an analytic function of the energy and spin angular momentum which are exact integrals of motion, and the vertical energy which is an approximate integral of motion. Thus, our model is not exactly in equilibrium. The alternative is to use angle-action variables \citep{binney2011,piffl2014,agama} but the computational overhead for doing so for the fitting procedure, where ${\cal O}(10^5)$ realizations of the model must be evaluated, is unfeasible. Moreover, the departures from equilibrium due to using the vertical energy are likely swamped by actual departures from equilibrium, as discussed above. In fact, we find angle-action variables extremely useful for analysing residuals of the best-fit model.

We first test our fitting procedure with mock data generated by \texttt{GalactICS}, a code designed to build equilibrium models for disk-bulge-halo systems \citep{Kuijken_Dubinski_1995, widrow2005, deg2019}. Stars for the mock sample are drawn from the same DF as is used in our statistical model. However, the potential in \textsc{GalactICS} is calculated self-consistently from the total mass distribution whereas the potential for our statistical model is a parametric function of $z$ and Galactocentric radius $R$. Thus, the mock data is not generated by the model, which, in some sense, makes it a stronger test of the method. Our analysis recovers the radial and vertical forces as well as low-order velocity moments of the DF. The fractional residuals in the potential, though only a few percent, are dominated by systematic errors. This reflects the fact that the potential for the mock data is not drawn from the model. On the other hand, the residuals for the DF are dominated by statistical errors.

We next fit a sample of $\sim$ 260 thousand giant stars from GDR2. Our results for the circular speed and vertical force in the Solar Neighborhood are consistent with literature values, though the rotation curve we measure is somewhat flatter. We examine residuals of star counts and velocity moments of DF in $R-z$ and $z-v_z$ phase-space planes. These residuals reveal hints of a coupling between in-plane and radial motions as well as the {\it Gaia} phase spirals discovered by \citet{antoja2018}. Finally, we use our best-fit model for the gravitational potential to transform star count residuals to action-frequency-angle coordinates. Here, we find sharp features corresponding to the {\it Gaia} phase spirals as well as velocity arches first seen in \citet{gaiadr2_vrad} and some of the moving groups identified by \citet{Trick2019}.

This paper is organized as follows: In Section \ref{AlgoIntro}, we present our model for the potential and DF as well as our method for computing and optimizing the likelihood function. In Section \ref{DataSel}, we describe our sample selection from GDR2. We test our fitting algorithm with mock data in Section \ref{result_Mock} and show the results from our GDR2 sample in Section \ref{result_Gaia}. In Section \ref{AAV} we show the star counts residuals in action-frequency-angle spaces. We discuss prospects for  improving the method in \ref{Discussion} and conclude with a summary of our results in Section \ref{Conclusion}.

\section{Preliminaries}\label{AlgoIntro}

In this section, we describe our models for the gravitational potential and stellar DF. We also outline our fitting procedure including the construction of the likelihood function. We work in Galactocentric cylindrical coordinates $(R,\phi,z)$, where $R$ is the in-plane distance to the Galactic Center, $\phi$ is the azimuthal angle towards the direction of Galactic rotation with the Sun at $\phi=0$, and $z$ is the displacement from the midplane. Correspondingly, we use $v_R$, $v_\phi$ and $v_z$ to denote radial, azimuthal and vertical velocities respectively. The distance to the Galactic Center is denoted by $r=\sqrt{R^2+z^2}$. We take the Solar Circle radius to be $R_0=8.3\,{\rm kpc}$ \citep{gillessen2009} and the Sun's displacement from the midplane to be $z_0=20.3\,{\rm pc}$ \citep{bennett2019}.

\subsection{Potential}\label{AlgoIntro-Potential}

We model the gravitational potential as the sum of contributions from the gas
and stellar disks, the bulge and the dark halo:
\begin{equation}\label{eq:totalpot}
    \Psi(R,z)=\Psi_g(z)+\Psi_b(r)+\Psi_d(R,z)+\Psi_h(r)~.
\end{equation}
We assume that the gas disk is razor thin and has constant surface density $\Sigma_g$. The potential is then given by
\begin{equation}\label{eq:gasdiskpot}
    \Psi_g(z)=2\uppi G\Sigma_g|z|
\end{equation}
where $G$ is the gravitational constant. As discussed in Section \ref{DataSel}, stars with $\left|R-R_0\right|>2\,{\rm kpc}$ and $\left|z-z_0\right|<80\,{\rm pc}$ are excluded from our sample. Therefore, the scale length and scale height of the gas disk are poorly constrained by the model. In effect, the potential in Equation \ref{eq:gasdiskpot} fixed the scale length to be infinity and the scale height to be zero. 

We take the contribution from the central bulge to be that of a point mass,
\begin{equation}\label{eq:Psi_bulge}
    \Psi_b(r)=-\frac{GM_b}r~,
\end{equation}
where $M_b$ is the bulge mass. That is, we assume that the mass of the bulge is entirely inside the Solar Circle and that its potential is well-approximated by the monopole term in a spherical harmonics expansion in our region of interest.

Next, we consider two possible forms for $\Psi_d(R,z)$: the Miyamoto-Nagai (MN) potential \citep{MNDisk}
\begin{equation}\label{eq:MNdisk}
    \Psi_{\rm MN}(R,z)=-\frac{GM_d}{\sqrt{R^2+\left(a+\sqrt{z^2+h^2}\right)^2}}
\end{equation}
and the potential $\Psi_{\rm ES}(R,z)$ of an exponential-sech-squared disk (ES), which satisfies the Poisson's equation for the density
\begin{equation}\label{eq:ESdisk}
    \rho_{\rm ES}(R,z)=\frac {M_d}{8\uppi a^2h}\exp {\left(-\frac{R}a\right)}
    {\rm sech}^2 \frac{z}{2h}~.
\end{equation}
In both models, $M_d$ is the disk mass and $a$ and $h$ are disk scale length and height respectively.

Finally, we adopt the NFW potential \citep{NFW} for the halo,
\begin{equation}\label{eq:Psi_halo}
    \Psi_h(r)=-\frac{4\uppi G\rho_0r_h^3}r\ln{\left(1+\frac r{r_h}\right)}~.
\end{equation}
The corresponding density profile is
\begin{equation}\label{eq:rho_halo}
    \rho_h(r)=\rho_0\left(\frac r{r_h}\right)^{-1}\left(1+\frac r{r_h}\right)^{-2}
\end{equation}
where $r_h$ is the halo scale radius and $\rho_0$ is a scale density. In this work, we have found it more convenient to parameterize the halo by the mass inside a sphere of radius $R_0$, i.e.
\begin{equation}\label{eq:Mhodot_def}
    M_{h\odot}\equiv\int_{0}^{R_0}4\pi r^2\rho_h(r)\dif r
\end{equation}
and the halo density at the position of the Sun, i.e.
\begin{equation}\label{eq:rhohodot_def}
\rho_{h\odot}\equiv \rho_h(R_0)
\end{equation}
The relationship between $(M_{h\odot},\rho_{h\odot})$ and $(\rho_0,r_h)$ can be easily obtained by combining Equations \ref{eq:Mhodot_def} and \ref{eq:rhohodot_def} to give
\begin{equation}\label{eq:SolveForrh}
    \frac{M_{h\odot}}{4\uppi{R_0}^3\rho_{h\odot}}
    =\varphi\left(\frac {r_h}{R_0}\right)
\end{equation}
where
\begin{equation}\label{eq:phix}
    \varphi(x)\equiv (x+1)^2\ln{\left(1+\frac 1x\right)}-x-1,\qquad x>0
\end{equation}
is a monotonically decreasing function with limits $\varphi(0^+)=+\infty$ and $\varphi(+\infty)=\frac 12$. The monotonicity of $\varphi(x)$ guarantees that we can numerically solve for $r_h$ from Equation \ref{eq:SolveForrh} and then solve for $\rho_0$ from the value of $M_{h\odot}$ or $\rho_{h\odot}$.

\subsection{Distribution function}\label{DF_sect}

As will be discussed in Section \ref{DataSel}, we use a sample of giant stars within $\sim2\,{\rm kpc}$ of the midplane. We assume that the stars come from a multi-component stellar disk that is in dynamical equilibrium and ignore the possibility that there are stars from the stellar halo or a stellar stream mixing into the sample. We build the model DF from analytic functions of the spin angular momentum $L_z=Rv_\phi$, the vertical energy,
\begin{equation}\label{eq:Ez}
    E_z\equiv\frac12 {v_z}^2+\Psi(R,z)-\Psi(R,0)~,
\end{equation}
and the in-plane energy
\begin{equation}\label{eq:Ep}
    E_p\equiv\frac 12\left(v_R^2+v_\phi^2\right)+\Psi(R,0)=E-E_z
\end{equation} 
(see \citet{Kuijken_Dubinski_1995} and references therein). The advantage of this approach is that for a given potential, the three integrals are explicit functions of the phase space coordinates and therefore are easy to compute. The main drawback is that while $E_z$ and $E_p$ are conserved to a good approximation for nearly circular orbits, they vary significantly along orbits that make large excursions in $R$ and $z$. Thus, a model that includes a warm disk will be somewhat out of equilibrium. We will have more to say about this issue in Section \ref{sec:lkfunc}.

We model the stellar DF as the sum of two terms,
\begin{equation}\label{eq:DF}
    f({\boldsymbol r},\,{\boldsymbol v})=\eta f_1({\boldsymbol r},\,\,{\boldsymbol v})+(1-\eta) f_2({\boldsymbol r},\,{\boldsymbol v})
\end{equation}
to account for thin and thick disk components. In this expression, the dimensionless constant $\eta\in(0,1)$ controls the relative contributions of the two components. For each disk, we write 
\begin{equation}\label{eq:DF_indv}
    f_i({\boldsymbol r},\,{\boldsymbol v})=\frac{\Omega}{\kappa}\frac{\rhot_i}{\sigmat_{Ri}^2\sigmat_{zi}}\exp{\left(-\frac{E_p-E_c}{\sigmat_{Ri}^2}-\frac{E_z}{\sigmat_{zi}^2}\right)}
\end{equation}
where $i=1,2$ \citep{Kuijken_Dubinski_1995}. Here, 
\begin{equation}
    E_c(R_c)=\Psi(R_c,0)+\frac12v_c^2
\end{equation}
is the energy of a particle on a circular orbit with angular momentum $L_z$ and $R_c$ is the guiding radius which depends implicitly on $L_z$. The angular frequency $\Omega$ and epicycle frequency $\kappa$ are well-known functions of $R_c$ (see e.g. \citealt{GalacticDynamics} and also Appendix \ref{app:norm}). By contrast, $\sigmat_{Ri}$, $\sigmat_{zi}$, and $\rhot_{di}$ are user-specified functions of $R_c$, which control the radial and vertical velocity dispersions and density in the midplane. Normally, one thinks of these quantities as functions of $R$. The essence of the \citep{Kuijken_Dubinski_1995} construction is to use $R_c$, which is an integral of motion, as a proxy for $R$, which isn't. In this work, we choose the following parametric forms for these quantities:
\begin{equation}
\begin{split}
    \rhot_{i}(R_c)\propto&\exp{\left(-\frac{R_c-R_0}{R_{\rho,i}}\right)}\\
    \sigmat_{Ri}(R_c)=&\sigma_{Ri,0}\exp{\left(-\frac{R_c-R_0}{R_{\sigma_R,i}}\right)}\\
    \sigmat_{zi}(R_c)=&\sigma_{zi,0}\exp{\left(-\frac{R_c-R_0}{R_{\sigma_z,i}}\right)}
\end{split}
\end{equation}
Note that the density scale parameters $\rho_{i,0}$ are absorbed into an overall normalization factor (see Section \ref{sec:lkfunc}) and $\eta$.

\subsection{Likelihood function and fitting procedure}\label{sec:lkfunc}

The likelihood function is given by the product of probabilities for individual stars,
\begin{equation}\label{eq:likelihood}
    \mathcal L=\prod_{i=1}^{N_s} \frac{f({\boldsymbol r}_i,\,{\boldsymbol v}_i)}{\mathcal N}
\end{equation}
where $N_s$ is the number of stars and 
\begin{equation}\label{eq:norm_form}
    \mathcal N\equiv\int f({\boldsymbol r},\,{\boldsymbol v})g({\boldsymbol r})\dif^3{\boldsymbol r}\dif^3{\boldsymbol v}
\end{equation}
is the normalization factor. The function $g({\boldsymbol r})$ accounts for our geometrical selection function as described in Section \ref{DataSel}. This form of the likelihood function has the attractive feature that it doesn't require binning. Details of the calculation of $\mathcal N$ are given in Appendix \ref{app:norm}.

As discussed in the introduction, exact equilibrium models can be constructed using action-angle variables rather than the explicit integrals of motion $(E_p,\,L_z,\,E_z)$ (see e.g. \citealt{binney2010, binney2011, piffl2014, Cole2017, agama}). However, the transformation between the usual phase space coordinates and action-angle variables requires its own set of approximations (see Section \ref{AAV_intro}). Moreover, the calculation of the normalization factor, which involves the geometric selection function $g(\boldsymbol r)$, would require a complicated and computationally expensive six-dimensional Monte Carlo integration. Since our fitting algorithm involves ${\cal O}(10^5)$ evaluations of the likelihood function, this approach seems unfeasible. By contrast, several of the integrals for the normalization factor can be done analytically when we work with 
$(E_p,\,L_z,\,E_z)$, as discussed in Appendix \ref{app:norm}.

In Table \ref{tab:prior} we list the model parameters. There are eighteen parameters in total: seven for the potential and eleven for the DF. Note that $a$, $h$, and $M_d$ have different meanings in the ES and MN disk potentials. We compute the potential and forces using the \texttt{Python} package \texttt{AGAMA} \citep{agama} and optimize the model over the model parameters via the Markov chain Monte Carlo (MCMC) sampler \textsc{emcee} \citep{emcee}. We adopt linear priors for all parameters as listed in Table \ref{tab:prior}. We also require that $M_{h\odot}>2\uppi R_0^3\rho_{h\odot}$ to ensure that Equation \ref{eq:SolveForrh} has a valid solution.
\begin{table}
    \centering
    \renewcommand{\arraystretch}{1.4}
    \begin{tabular}{cl}
    Parameter &  Prior range \\
    \hline
    $\Sigma_g$ & $\left[0,\,20\right]\times 10^7\,{\rm M_\odot/kpc^2}$\\
    $M_b$ & $\left[0,\,50\right]\times 10^9\,{\rm M_\odot}$ \\
    $M_d$ & $\left[0.5,\,20\right]\times10^{10}\,{\rm M_\odot}$ \\
    $a$ & $\left[0.5,\,10\right]\,{\rm kpc}$ \\
    $h$ & $\left[0,\,2\right]\,{\rm kpc}$ \\
    $M_{h\odot}$ & $\left[0.1,\,10\right]\times10^{10}\,{\rm M_\odot}$ \\
    $\rho_{h\odot}$ & $\left[1,\,100\right]\times10^{6}\,{\rm M_\odot/kpc^3}$ \\
    \hdashline
    $R_{\rho1}$ and $R_{\rho2}$ & $\left[0.5,\,20\right]\,{\rm kpc}$\\
    $\sigma_{R1,0}$ and $\sigma_{R2,0}$ &$\left[5,\,100\right]\,{\rm km/s}$\\
    $R_{\sigma_R,1}$ and $R_{\sigma_R,2}$ & $\left[0.5,\,50\right]\,{\rm kpc}$\\
    $\sigma_{z1,0}$ and $\sigma_{z2,0}$ & $\left[1,\,100\right]\,{\rm km/s}$\\
    $R_{\sigma_z,1}$ and $R_{\sigma_z,2}$ & $\left[0.5,\,50\right]\,{\rm kpc}$\\
    $\eta$ & $\left[0.5,1\,\right]$\\
    \end{tabular}
    \caption{Parameter prior ranges adopted in this work.}
    \label{tab:prior}
\end{table}

\subsection{Moments of distribution function}\label{moment_sect}

As discussed above, one of our goals is to examine residuals of the best-fit model since these may reveal manifestations of departures of the disk from equilibrium. In general, it is unfeasible to do this in the full six-dimensional phase space since residuals generally require some sort of binning. We therefore set up the machinery to examine moments of the DF in the $R-z$ space or the meridonal plane (subscript ``mer") and the $z-v_z$ space or the vertical phase space plane (subscript ``ver"). For example, the number densities in these two planes are given by
\begin{equation}\label{eq:nRz}
    n_{\rm mer}(R,z)=\mathcal N_{\rm mer}\int f({\boldsymbol r},\,{\boldsymbol v})g({\boldsymbol r})R\dif\phi\dif^3{\boldsymbol v}
\end{equation}
and
\begin{equation}\label{eq:nperp}
    n_{\rm ver}(z,v_z)=\mathcal N_{\rm ver}\int f({\boldsymbol r},\,{\boldsymbol v})g({\boldsymbol r})R\dif\phi\dif R\dif v_R\dif v_\phi
\end{equation}
where ${\cal N}_{\rm mer}$ and ${\cal N}_{\rm ver}$ are normalization factors such that the model-predicted total number counts matches the data. Likewise, the mean azimuthal velocities in these planes are given by
\begin{equation}\label{eq:vphimean_Rz}
    \langle v_\phi\rangle_{\rm mer}(R,\,z)
    =\frac{\mathcal N_{\rm mer}}{n_{\rm mer}(R,z)}\int v_\phi f({\boldsymbol r},\,{\boldsymbol v})(R\dif\phi)\dif^3{\boldsymbol v}
\end{equation}
and
\begin{equation}\label{eq:vphiperp}
    \langle v_\phi\rangle_{\rm ver}(z,v_z)
    =\frac{\mathcal N_{\rm ver}}{n_{\rm ver}(z,v_z)}
    \int v_\phi f({\boldsymbol r},\,{\boldsymbol v})g({\boldsymbol r})R\dif R\dif\phi\dif v_R\dif v_\phi\\
\end{equation}
By symmetry, the model predicts $\langle v_R\rangle = 0$ in both spaces. Finally, we note that $n_{\rm ver}$, $\langle v_R\rangle_{\rm ver}$ and $\langle v_\phi\rangle_{\rm ver}$ correspond to the three views of the {\it Gaia} phase spirals in \citet{antoja2018}. The calculation of these moments is discussed in Appendix \ref{app:dens_prof}.

\section{Data Selection}\label{DataSel}

In this section, we describe our sample selection. As in Paper I we draw our sample from the catalog {\it gaiaRVdelpeqspdelsp43}\footnote{See https://zenodo.org/record/2557803 for their data} \citep{sch_prlx}. This catalog includes stars in GDR2 with complete 6D phase space measurements and corrects for systematic biases in {\it Gaia}'s parallaxes. We use {\it E\_dist} from the catalog as the distance $d$ from the Sun and take
\begin{equation}
\varepsilon_d=\sqrt{distm2-E\_dist^2}
\end{equation}
to be its uncertainty, where {\it distm2} is the expectation value of $d^2$. To ensure precision in parallax measurements, we implement the same quality cuts as in Paper I, which mainly follow those recommended by \citet{sch_prlx}:
\begin{itemize}
\item Photometry: $3<G<14.5$, $G_{RP}>0$ and $G_{BP}>0$
\item Radial velocity uncertainty: $\error_{v_{\rm rad}}<10\,{\rm km/s}$
\item Parallax uncertainty: $\error_\varpi<0.1\,{\rm mas}$ and $\varpi/\error_\varpi > 5$
\item Visibility period: $n_{\rm vis}>5$
\item BP/RP flux excess factor range: $1.172<\verb!bp_rp_excess_factor!<1.3$
\item Minimum heliocentric distance constraint: $d>80\,{\rm pc}$
\end{itemize}
As discussed in Paper I, we guarantee completeness by selecting stars from a particular region of the color-magnitude diagram. The completeness constraint comes mainly from the availability of radial velocity measurements, which are based on stellar spectra and generally available only for the brightest stars. We therefore add the following photometry cuts to select giant stars:
\begin{itemize}
    \item $B$ minus $R$ color: $G_{BP}-G_{RP}>1$
    \item Absolute $G$-band magnitude: $M_G<2$
    \end{itemize}
Finally, we remove stars with Galactocentric speed greater than $550\,{\rm km/s}$ which is the approximate escape speed of the Galaxy \citep{escape_vel}. We also remove stars identified by \citet{vradcontam} as potentially having large radial velocity errors due to contamination of their spectra by their neighbors\footnote{See https://arxiv.org/src/1901.10460v1/anc/ for a catalog of these stars}. 

We calculate the positions and velocities of stars in our sample using the \texttt{astropy.coordinates} Python package \citep{astropy1,astropy2}. We take the Sun's peculiar velocities to be $(U_\odot,\,V_\odot,\,W_\odot)=(11.1,\,12.24,\,7.25)\,{\rm km/s}$ \citep{schonrich2010} and the rotation speed at Solar Circle to be $v_{c\odot}=220\,{\rm km/s}$ \citep{MWPot2014} in the calculation.

To avoid problems with extinction, we exclude stars within $80\,{\rm pc}$ of the midplane and those with Galactic latitude $|b|<15^\circ$ \footnote{We measure $b$ with regard to the Galactic midplane, instead of the direction from the Sun pointing towards the Galactic Center.}. In addition, we include only those stars in our local patch of the disk by requiring that $|\phi|<4^\circ$, $|R-R_0|<2\,{\rm kpc}$ and $|z-z_0|<2\,{\rm kpc}$. Heliocentric distances within our sample volume $\mathcal{V}$ range from $r_{\rm min}=0.08\,{\rm kpc}$ to $r_{\rm max}=2.901\,{\rm kpc}$ from the Sun. Thus, given our apparent magnitude cut $3<G<14.5$, we require an absolute magnitude cut of
\begin{equation}
\begin{split}
    3-5{\rm log_{10}}\frac{r_{\rm min}}{10\,{\rm pc}}&=-1.52<M_G<\\
    14.5-5{\rm log_{10}}\frac{r_{\rm max}}{10\,{\rm pc}}&=2.19
\end{split}
\end{equation}
to avoid the Malmquist bias. If we combine these cuts with the constraint $M_G<2$ for giants, we arrive at an absolute magnitude cut of $-1.52<M_G<2$, which is the same as was used Paper I. The geometrical selection function $g(\boldsymbol r)$ is simply the Heaviside function: unity inside ${\cal V}$ and zero outside. The final sample has $\sim260$ thousand stars.

\section{Mock data}\label{result_Mock}

In this section, we describe tests of our method based on mock data. The data are created using the code \texttt{GalactICS}, which was designed to generate equilibrium initial conditions for N-body simulations of isolated disk galaxies \citep{Kuijken_Dubinski_1995, widrow2005, deg2019}. The DF in \texttt{GalactICS} is given by Equation \ref{eq:DF_indv}, which is the same as the DF in our statistical model. However, the potential in \texttt{GalactICS} is calculated from the total density via Poisson's equation whereas our statistical model uses a parametric expression for the potential. The working assumption is that this parametric expression is flexible enough to accurately model the ``true" potential (i.e, the potential used to generate the mock data) but there is no {\it a priori} guarantee of this. As we will see, this leads to systematic errors in the recovery of the potential and force. In principle, these errors can be reduced by choosing a more general parametric model for the potential. Of course, we expect similar systematic errors in GDR2, which is why we perform the mock tests in this way.

\subsection{mock data sample}\label{Mock_contour}

We begin by constructing a self-consistent \texttt{GalactICS} model that comprises a thin disk, a thick disk, a bulge, and a dark halo. The model is chosen to qualitatively match characteristics of the Milky Way. In particular, the thin disk has a mass of $3.7\times 10^{10}\,M_\odot$, a radial scale length of $2.5\,{\rm kpc}$ and a vertical scale height of $300\,{\rm pc}$. The corresponding values for the thick disk are $1.2\times 10^{10}\,M_\odot$, $3.5\,{\rm kpc}$, and $900\,{\rm pc}$. We sample the DFs for the two disks with the same geometric cuts that will be applied to the real data and arrive at a final sample size of 250,730, which is very close to that of our GDR2 sample.

\subsection{parameter estimation}\label{sec:mockparam}

\begin{figure}
	\includegraphics[width=\columnwidth]{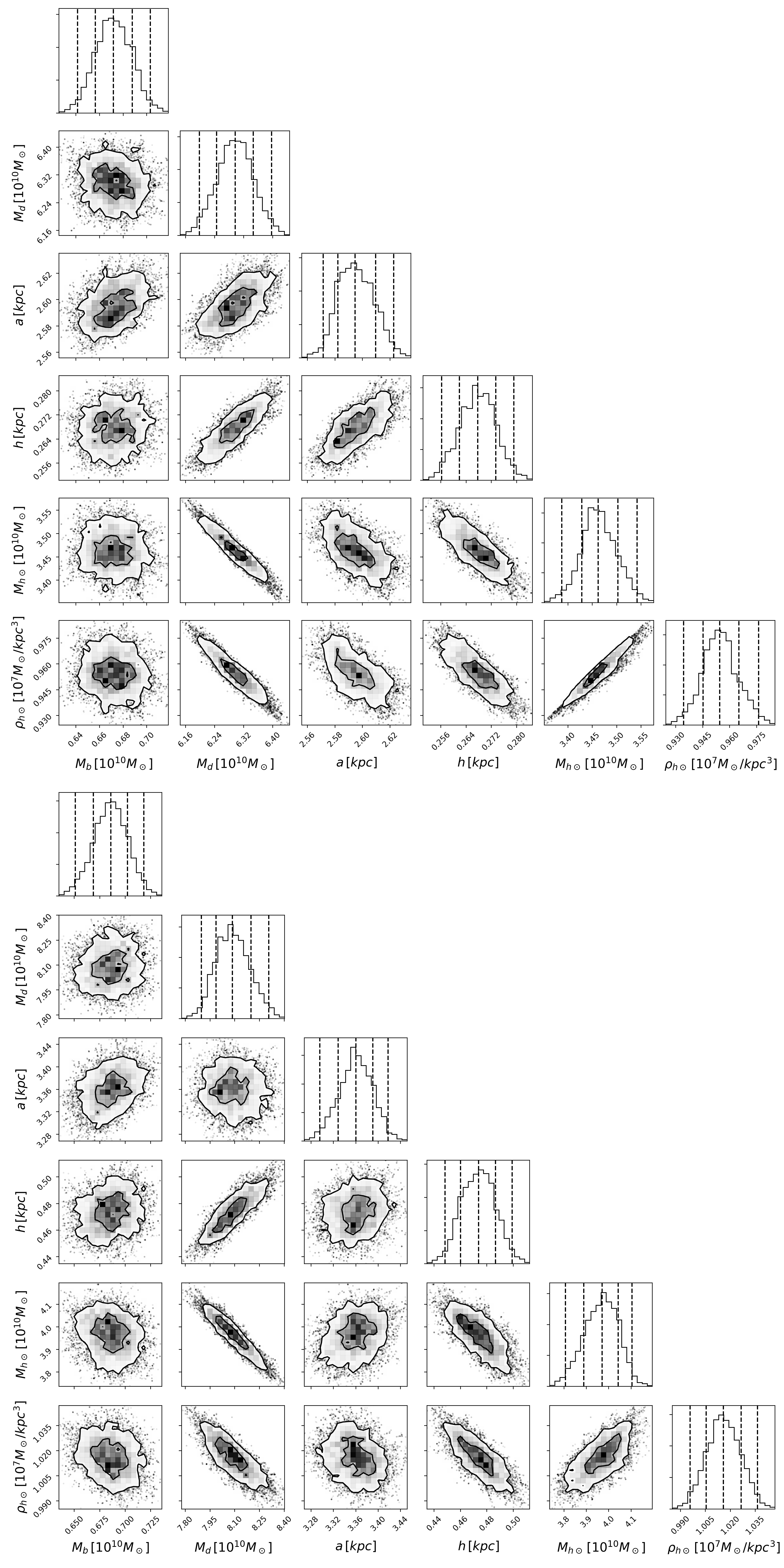}
    \caption{2D projections of the posterior PDF for the model parameters with ES (top) and MN (bottom) disks. We only show projections for the six model parameters associated with the gravitational potential excluding $\Sigma_g$.}
    \label{fig:MockContour}
\end{figure}

We use the MCMC sampler \texttt{emcee} to map out the {\it posterior} probability distribution function (PDF) of the model parameters. Our choice of starting values for the walkers is guided by global properties of the Milky Way. The chain burns in quickly, as can be verified by inspecting values of each parameter as a function of position along the chain. In Figure \ref{fig:MockContour} we show two dimensional projections of PDFs for the six parameters associated with the potential excluding $\Sigma_g$, because we set $\Sigma_g=0$ as there is no gas disk in the \texttt{GalactICS} model. We include results from both ES and MN models for comparison. These figures reveal a number of strong correlations between various pairs of parameters. In particular, there is a positive correlation between $M_d$ and $h$ and negative correlations between $M_d$ and both $M_{h\odot}$ and $\rho_{h\odot}$. The $M_d-M_{h\odot}$ correlation works to keep the total mass interior to the Sun constant. On the other hand, the $M_d-h$ correlation works to keep the total density in the Solar Neighborhood roughly constant since an increase in the disk thickness can be compensated by an increase in the disk mass. The $M_d-\rho_{h\odot}$ correlation also works to keep the density in the Solar Neighborhood roughly constant. Note that this correlation is much tighter for the ES disk than the MN one. The difference is likely due to the difference in the vertical structure between the two models. In the ES model, the mass distribution is more tightly confined to the plane, with an exponential rather than power-law fall off with increasing $|z|$. The conclusion from these entirely expected correlations, is that the data constrain locally rather than globally defined quantities from the potential. Inspection of 2D projections for the full 18-dimensional parameter space didn't reveal any further strong correlations.

\subsection{rotation curve and gravitational force}\label{Mock_RotCurve}

\begin{figure}
	\includegraphics[width=\columnwidth]{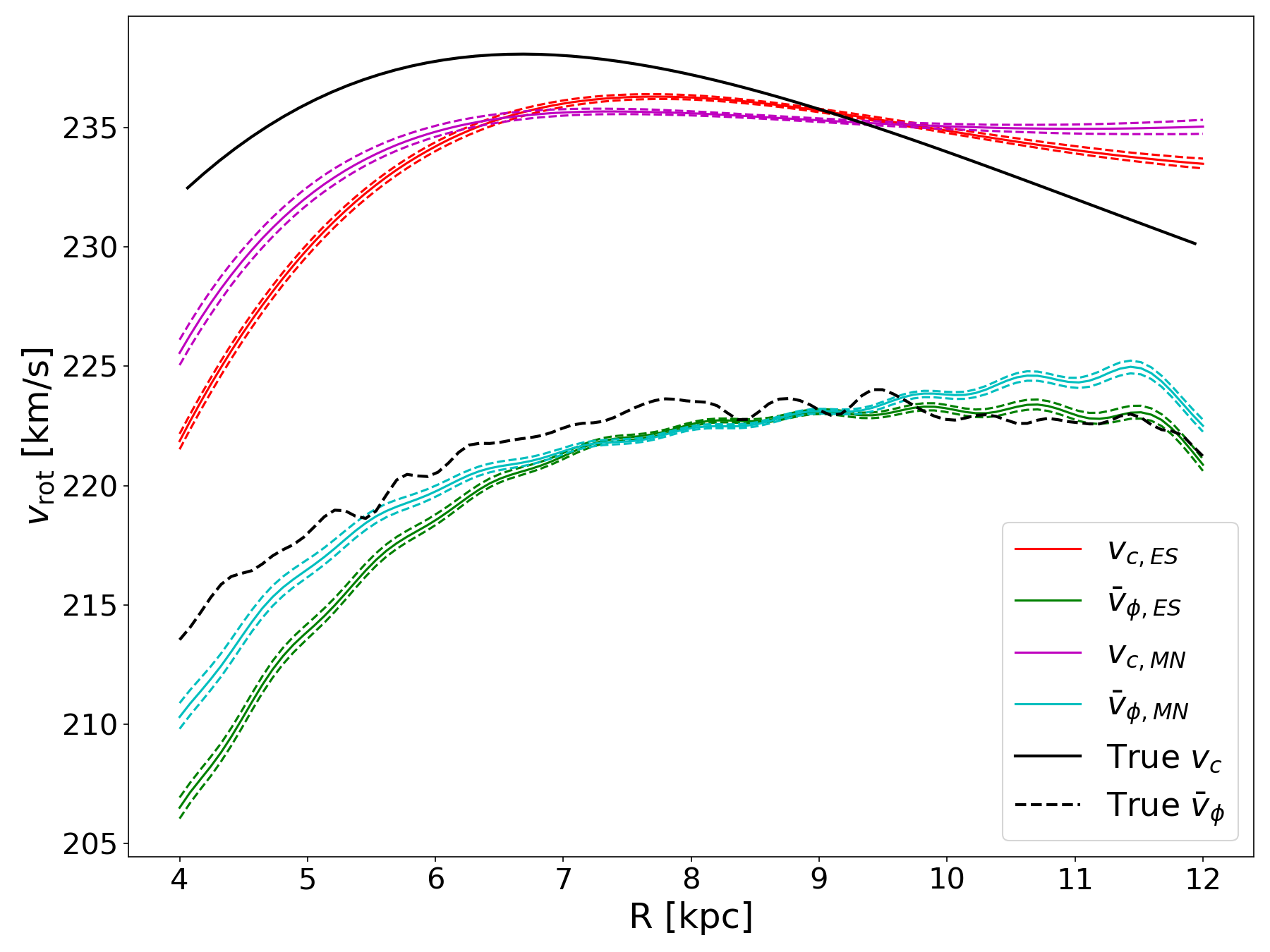}
    \caption{Model predictions and true values for the rotation curve. We show the predictions for the circular speed curve as derived from the gravitation potential for the ES (red) and MN (purple) models as well as the true circular speed curve (black). The $1\sigma$ uncertainties are shown as dashed lines. We also show model predictions for the rotation curve $\overline v_\phi$ for the ES (green) and MN (cyan) models along with the rotation curve derived directly from the mock data (black dotted curve).}
    \label{fig:MockRotCurve}
\end{figure}

In Figure \ref{fig:MockRotCurve}, we plot model predictions for the circular speed curve $v_{\rm circ}$ as determined from the best-fit potential, and the rotation curve $\overline v_\phi(R)$ as determined from the best-fit DF via Equation \ref{eq:vphimean_Rz} with $z=0$. We also show the true $v_{\rm circ}$, which is derived directly from the potential in \texttt{GalactICS} and the true $\overline v_\phi(R)$, which is determined by computing the average azimuthal velocity of mock stars within $50\,{\rm pc}$ of the midplane. The difference between $v_{\rm circ}$ and $\overline v_\phi(R)$ is due to asymmetric drift (see e.g. \citealt{GalacticDynamics}). Our model correctly accounts for this difference and recovers both curves to within $\sim 1\% - 2\%$ between $6\,{\rm kpc}$ and $10\,{\rm kpc}$. Note that the ES and MN models agree remarkably well in their predictions for $\overline v_\phi(R)$ over this range, which is not surprising since the mean azimuthal velocity near the Sun is directly reflected in the data. Figure \ref{fig:MockRotCurve} also shows 1$\sigma$ error ranges that are estimated by sampling potential parameters from the MCMC chain. Overall, these error bars under-predict differences between the model and the mock data. Indeed, the residuals for the rotation curve are comparable to the difference between ES and MN predictions. Thus, we can use the difference between predictions by these two models as an estimate of systematic errors.

\begin{figure}
	\includegraphics[width=\columnwidth]{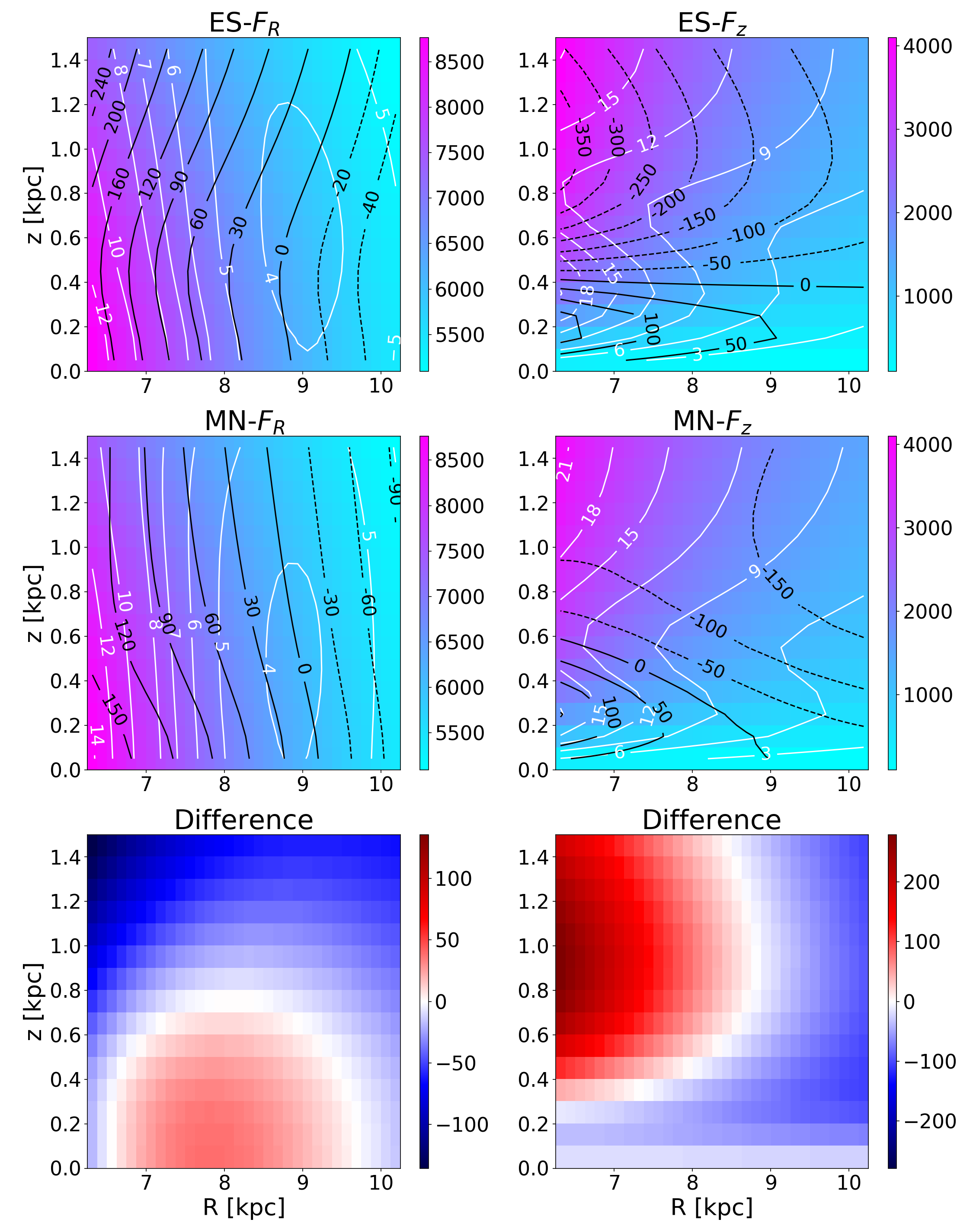}
    \caption{Inferred force from the analysis of mock data for both ES model (top row) and MN model (middle row). We show results of the radial and vertical components (left and right columns, respectively) as a color map in units of ${\rm (km/s)^2/kpc}$. We also show $1\sigma$ statistical uncertainties from the model (white contours) and the true$-$model residuals (black contours). The bottom row shows the difference between the two models.}
    \label{fig:Mock_F}
\end{figure}

This last point is further illustrated in Figure \ref{fig:Mock_F} where we compare predictions from the ES and MN models for the vertical and radial forces. The estimated statistical errors are a factor of $10\sim20$ times smaller than the residuals between either model and the true force, but are comparable to the difference between the models themselves.

\subsection{Distribution function and its moments}\label{Moment_Mock}

\begin{figure}
	\includegraphics[width=\columnwidth]{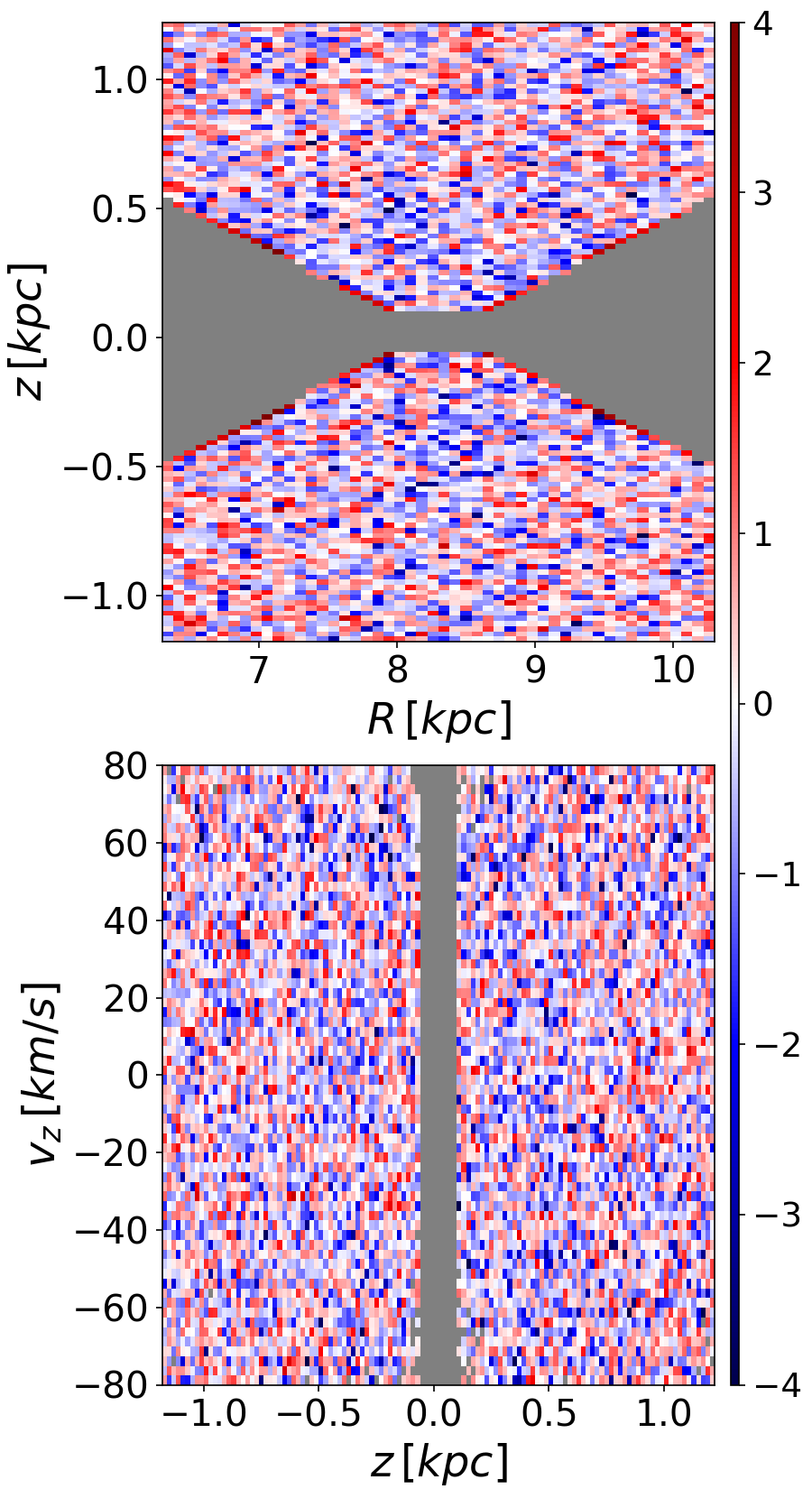}
    \caption{Data $-$ Model residuals in meridonal and vertical phase space planes scaled by the square root of the star counts in each pixel from data for the $R-z$. The bin size in the meridonal plane is $80\,{\rm pc}\times20\,{\rm pc}$ while the bin size in the $z-v_z$ is $20\,{\rm pc}\times2.5\,{\rm km/s}$). We only show results for the ES model, as the results for the MN model are hardly visually distinguishable.}
    \label{fig:Mock_Dens}
\end{figure}

\begin{figure}
	\includegraphics[width=\columnwidth]{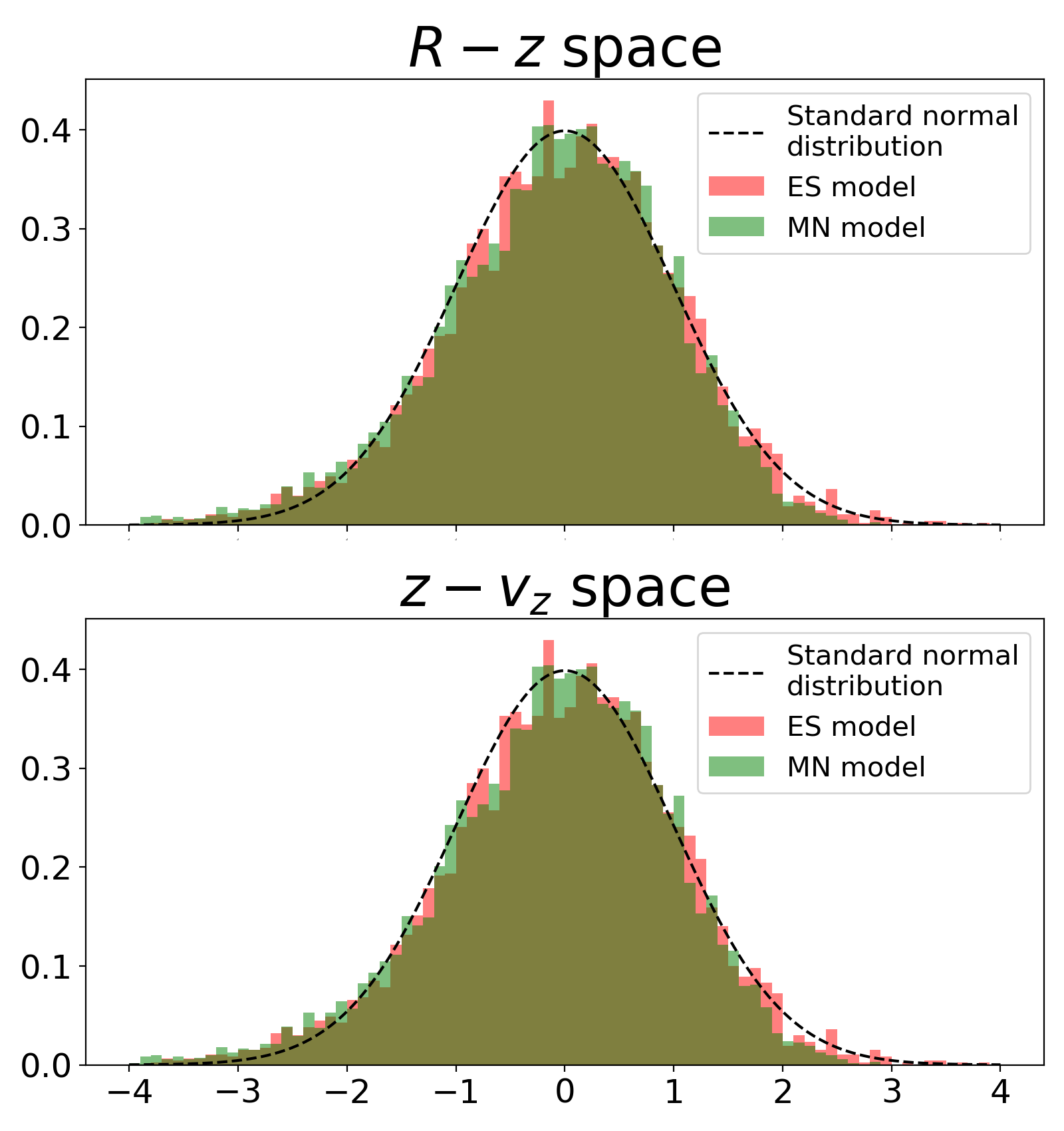}
    \caption{Probability density as derived from histograms of $\tilde\sigma_n$ for the $R-z$ and $z-v_z$ phase space projections. The black dashed line shows the standard normal distribution.}
    \label{fig:Mock_DensHist}
\end{figure}

As a further test of our fitting procedure, we compare model predictions for the number counts and velocity moments in the meridonal and vertical phase space planes. We first consider the residuals of the number counts scaled by the square root of star counts in each pixel in either plane. That is, we compute
\begin{equation}
    \tilde\sigma_n = \frac{n_d - n_m}{\sqrt{n_d}}
\end{equation}
where $n_d$ is the number of stars in a given pixel and $n_m$ is the model prediction for $n_d$ as computed in the center of that pixel using Equation \ref{eq:nRz} or \ref{eq:nperp}. The scaled residuals are shown in Figure \ref{fig:Mock_Dens} and appear to be random with no evidence for systematic errors that depend on $R$, $z$, or $v_z$. In Figure \ref{fig:Mock_DensHist}, we show the probability densities of $\tilde\sigma_n$ for the $R-z$ and $z-v_z$ planes and the ES and MN models. We find that they are all well-approximated by the standard normal distribution, which lends credence to the contention that the statistical errors from the MCMC analysis properly account for errors in the model.

Finally, we consider residuals of $\langle v_R\rangle_{\rm ver}$ and $\langle v_\phi\rangle_{\rm ver}$ in the vertical phase space plane. These plots are analogous to those shown by \citet{antoja2018} in their discovery paper of the {\it Gaia} phase spirals. We will show similar versions of these plots when we turn to GDR2 data. Note that the model prediction for $\langle v_R\rangle_{\rm ver}$ is zero and therefore the residual of $\langle v_R\rangle_{\rm ver}$ is just the data value. The situation with $\langle v_\phi\rangle_{\rm ver}$ is more complicated since it involves the rotation of the disk and asymmetric drift. Furthermore, we expect $\langle v_\phi\rangle_{\rm ver}$ to depend on the vertical and in-plane energies since $\sigma_z$ and $\sigma_R$ both depend on $L_z$, as seen in Equation \ref{eq:DF_indv}. The importance of this coupling for the nature of the phase spiral was stressed by \citet{Binney2018} and \citet{Darling2019}.

\begin{figure}
	\includegraphics[width=\columnwidth]{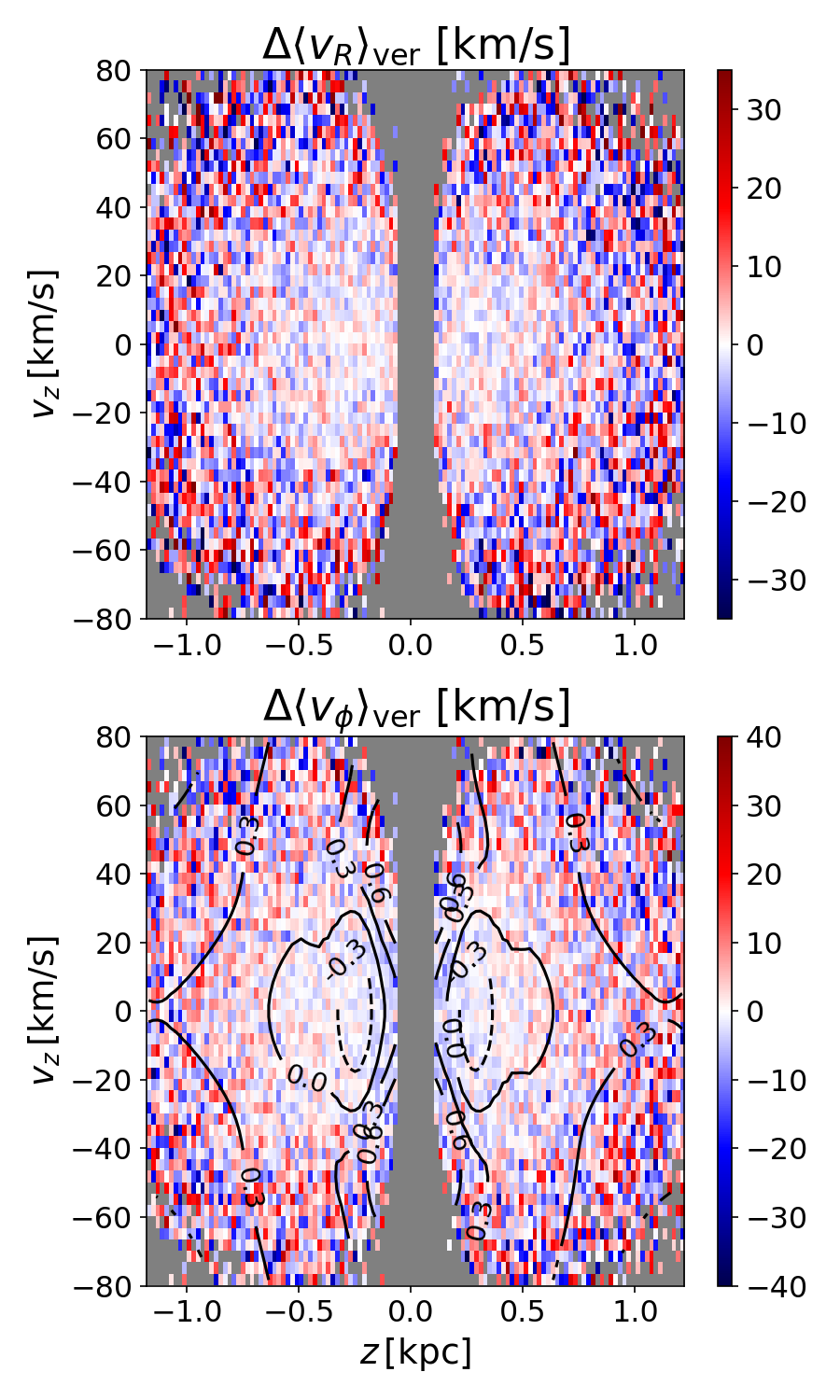}
    \caption{Map of residuals for the mean radial velocity (top panel) and mean azimuthal velocity (bottom panel) in the $z-v_z$ plane. Results are shown for the ES model. In the bottom panel, the difference between model predictions for the MN and ES models is indicated by the black contours. Pixels with fewer than 10 stars are colored grey.}
    \label{fig:Mock_AntojaSpiral}
\end{figure}

Maps of the residuals for $\langle v_R\rangle_{\rm ver}$ and $\langle v_\phi\rangle_{\rm ver}$ are shown in Figure \ref{fig:Mock_AntojaSpiral}. The residuals appear to be randomly distributed with zero mean. The increase in the RMS of the residuals as one moves out from the origin comes from the decrease in the number of particles per $z-v_z$ pixel. In the $\Delta\langle v_\phi\rangle_{\rm ver}$ panel, we plot the residuals for the ES model and overlay contours that show the difference between the ES and MN models \footnote{In this work, all differences between ES and MN model predicted value are defined as ES predictions minus MN predictions.}. Evidently, the two models make nearly identical predictions for $\langle v_\phi\rangle_{\rm ver}$. In summary, the model does an excellent job of recovering the DF.

\section{Results on GDR2 sample fitting}\label{result_Gaia}

In this section, we present our results for the GDR2 data. The procedure is the same as was used in our analysis of mock data except that here, the data are presented in celestial coordinates $(\alpha,\delta,\varpi,v_{\rm los},\mu_\alpha^*=\dot\alpha\cos\delta,\mu_\delta=\dot\delta)$. To account for uncertainties in these coordinates, we generate ten data sets that add in random errors under the assumption that the quoted errors are Gaussian. For each of these data sets, we convert celestial coordinates to positions and velocities and apply the selection criteria as described in Section \ref{DataSel}. Note that the data sets end up with slightly different numbers of stars. We find that the mean sample size is 264,150 stars with a standard deviation of 124. All results in this section are derived from the combined parameter chains for the ten data sets.

\subsection{parameter estimation}

\begin{figure}
	\includegraphics[width=\columnwidth]{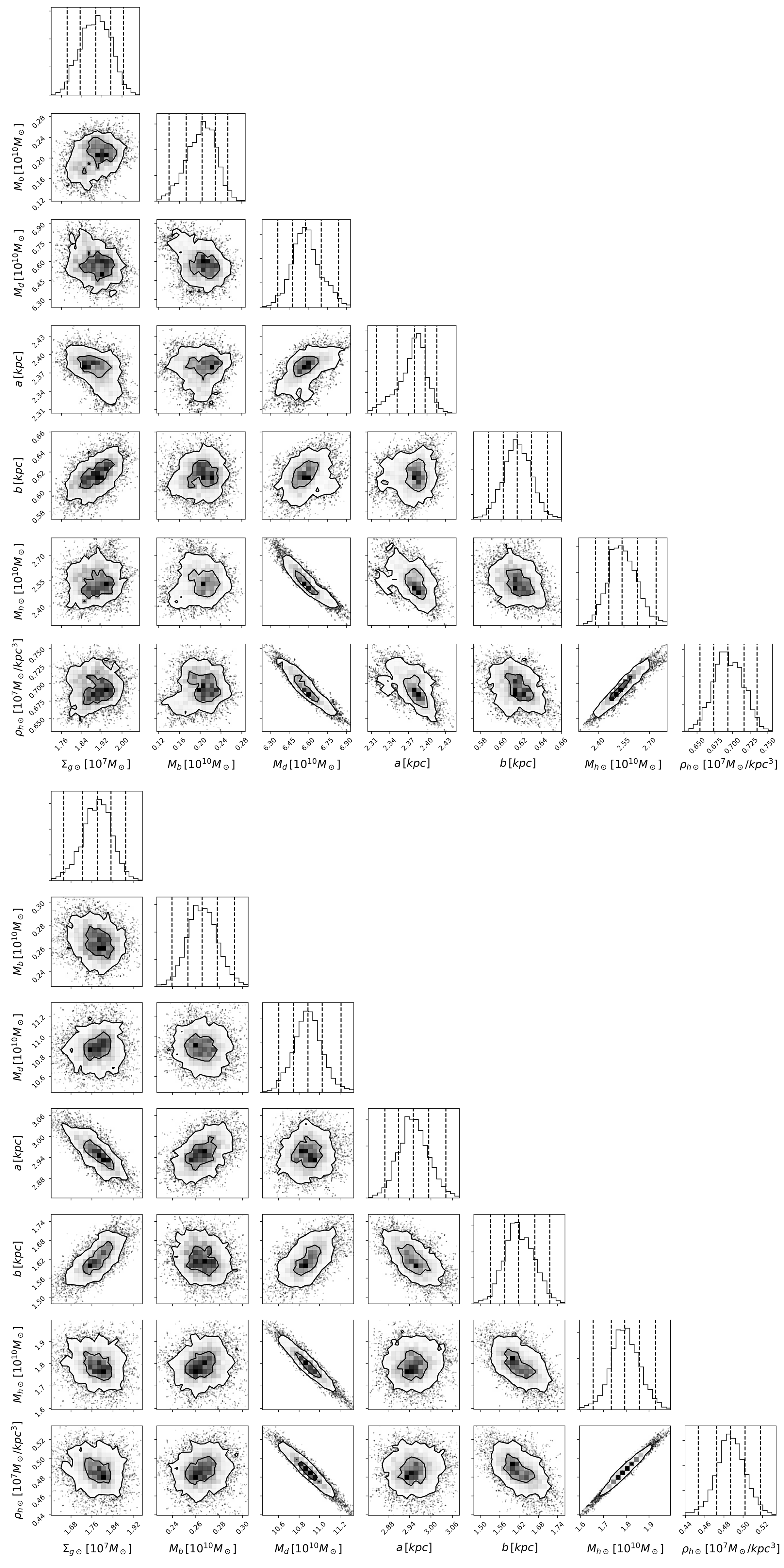}
    \caption{2D projections of the posterior PDF for parameters from the ES model (top) and MN model (bottom). Similar as Figure \ref{fig:MockContour}, we only show results for parameters associated with the potential.}
    \label{fig:GaiaContour}
\end{figure}

For the first data set, we use 50 walkers and require 5000 steps before convergence is reached. For other nine data sets, the burn-in period is shorter at around 1000 steps since we are able to use results from the first data set as a starting point. In Figure \ref{fig:GaiaContour} we show two-dimensional projections of the posterior PDF for the potential parameters. The PDF is qualitatively similar to the one we found for the mock data. We do not find any strong correlations among the parameters for the DF or between the DF and potential parameters. The three strong correlations that have been discussed in Section \ref{sec:mockparam} also appear in Figure \ref{fig:GaiaContour}. In addition, there is a weak negative correlation between $\Sigma_g$ and $a$ and a positive correlation between $\Sigma_g$ and $h$. Recall that our gas disk is assumed to be razor thin and have constant surface density in $R$. Thus, an increase in $\Sigma_g$ can be compensated by a decrease in $a$ and/or increase in $h$.

The best-fit parameters and $1\sigma$ uncertainties for our two models are presented in Table \ref{tab:GDR2Result}. The predictions for the DF parameters from the ES and MN models are strikingly similar. Evidently, the data tightly constrain the stellar DF. The differences are more pronounced for the potential, which is perhaps not surprising since they assume very different functional forms for the disk contribution. We should stress that the model is only sensitive to the total potential and that the gas-disk-bulge-halo decomposition in Equation \ref{eq:totalpot} should be interpreted cautiously. Nevertheless, we can compare our result for the gas disk with literature values. For example, \citet{Flynn2006} proposed $\Sigma_g=13.2{\rm M_\odot/pc^2}$, which was adopted by \citet{Zhang2013, Xia2016, Guo2020} as a fixed parameter. In this work, we obtain values that are higher by $35\%\sim45\%$.
\begin{table}
    \centering
    \renewcommand{\arraystretch}{1.4}
    \begin{tabular}{lll}
    Parameter and unit & ES-model & MN-model \\
    \hline
    $\Sigma_g\,\left[{\rm M_\odot/pc^2}\right]$ & $19.0\pm0.6$ & $17.8_{-0.5}^{+0.6}$\\
    $M_b\,\left[10^9\,{\rm M_\odot}\right]$ & $2.04_{-0.26}^{+0.31}$ & $2.65\pm0.01$\\
    $M_d\,\left[10^{10}\,{\rm M_\odot}\right]$ & $6.58_{-0.13}^{+0.10}$ & $10.90\pm0.14$\\
    $a\,\left[{\rm kpc}\right]$ & $2.38_{-0.02}^{+0.03}$ & $2.95\pm0.04$\\
    $b\,\left[{\rm kpc}\right]$ & $0.616\pm0.014$ & $1.62_{-0.05}^{+0.04}$\\
    $M_{h\odot}\,\left[10^{10}\,{\rm M_\odot}\right]$ & $2.54_{-0.09}^{+0.08}$ & $1.79\pm0.06$\\
    $\rho_{h\odot}\,\left[10^{-3}\,{\rm M_\odot/pc^3}\right]$ & $6.93_{-0.23}^{+0.20}$ & $4.87\pm0.15$\\
    $R_{\rho1}\,\left[{\rm kpc}\right]$ & $3.59_{-0.12}^{+0.15}$ & $3.91_{-0.16}^{+0.13}$\\
    $\sigma_{R_0,1}\,\left[{\rm km/s}\right]$ & $32.3\pm0.1$ & $32.4\pm0.1$\\
    $R_{\sigma_R,1}\,\left[{\rm kpc}\right]$ & $10.9\pm0.2$ & $11.2\pm0.2$\\
    $\sigma_{z_0,1}\,\left[{\rm km/s}\right]$ & $15.5\pm0.1$ & $15.4\pm0.1$\\
    $R_{\sigma_z,1}\,\left[{\rm kpc}\right]$ & $14.1_{-0.5}^{+0.4}$ & $14.8_{-0.5}^{+0.4}$\\
    $\eta$ & $0.864\pm0.003$ & $0.858\pm0.003$\\
    $R_{\rho2}\,\left[{\rm kpc}\right]$ & $4.28_{-0.11}^{+0.14}$ & $4.76_{-0.15}^{+0.16}$\\
    $\sigma_{R_0,2}\,\left[{\rm km/s}\right]$ & $43.8\pm0.2$ & $44.0\pm0.2$\\
    $R_{\sigma_R,2}\,\left[{\rm kpc}\right]$ & $8.32\pm0.08$ & $8.36\pm0.05$\\
    $\sigma_{z_0,2}\,\left[{\rm km/s}\right]$ & $30.0\pm0.1$ & $30.0\pm0.1$\\
    $R_{\sigma_z,2}\,\left[{\rm kpc}\right]$ & $8.25_{-0.14}^{+0.12}$ & $8.42_{-0.10}^{+0.09}$
    \end{tabular}
    \caption{Best-fit values and $1\sigma$ uncertainties from the fitting of the GDR2 sample.}
    \label{tab:GDR2Result}
\end{table}

\subsection{rotation curve and the Oort constants}\label{sect:gaia_rotcurve}

\begin{figure}
	\includegraphics[width=\columnwidth]{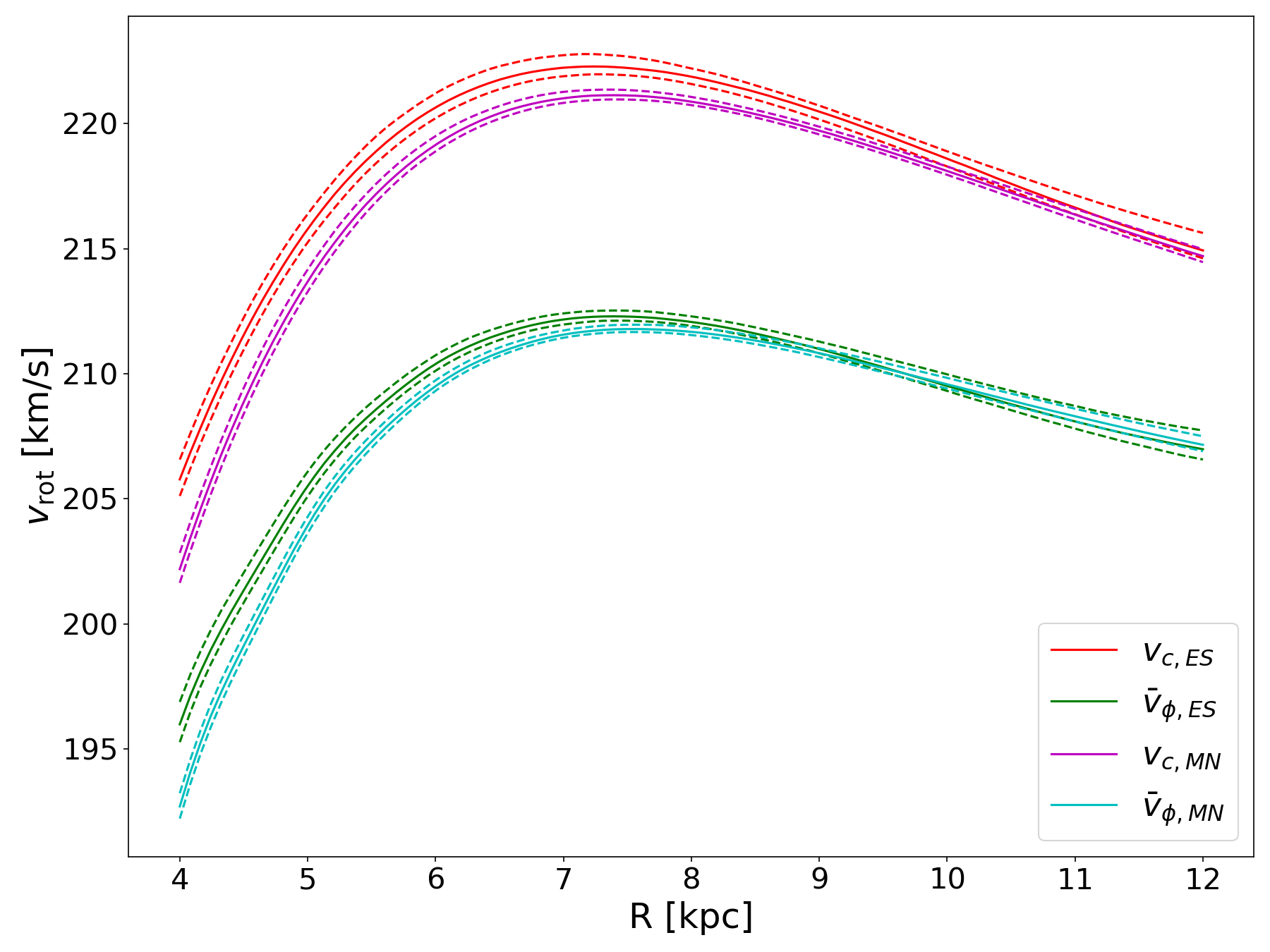}
    \caption{Predictions for the circular speed and rotation curve. Line types are the same as in Figure \ref{fig:MockRotCurve}.}
    \label{fig:GaiaRotCurve}
\end{figure}

In Figure \ref{fig:GaiaRotCurve}, we show model predictions for $v_{\rm circ}$ and $\overline v_\phi$ in the midplane for $4\,{\rm kpc} < R< 12\,{\rm kpc}$. Predictions for $v_{\rm circ}$ and $\overline v_\phi$ at the position of the Sun are given in Table \ref{tab:Gaia_rot_curve}. As we saw with the mock data test, the formal statistical uncertainties for the rotation curve are extremely small. Individually, the ES and MN models do not account for systematic errors that arise because they assume restricted functional forms for the potential. In short, we are over-fitting the data. The consistency of the two models in their predictions of $\overline v_\phi$ near the Sun indicates that this is the most secure prediction. Indeed, the predictions of $\overline v_\phi$ at the Sun from the two models are within the statistical uncertainties. On the other hand, the ES and MN predictions for $v_{\rm circ}$ at the Solar Circle differ by about a factor of five over the formal $1\sigma$ statistical uncertainties. We take that difference to be an indication of the systematic uncertainties in the model. We stress that this difference is only $\sim1\,{\rm km/s}$, and both model predictions are consistent with the literature value $219.1\pm14.1\,{\rm km/s}$ as the average of measurements in \citet{Vityazev2017, Bobylev2018, Krisanova2020} and \citet{Nouh2020}.

Our analysis of the GDR2 data indicates an asymmetric drift of $v_a\simeq 10\,{\rm km/s}$. The mean azimuthal velocity of stars satisfying $|R-R_0|<50\,{\rm pc}$ and $|z-z_\odot|<100\,{\rm pc}$ is $212.3\,{\rm km/s}$, which agrees very well with the local measurement of $\overline v_\phi$. We find that $\left<v_R^2\right>=1217\,{\rm (km/s)^2}$. Following the arguments in Section 4.8.2 of \citet{GalacticDynamics}, we find $v_a = \left<v_R^2\right>/\left (80\,{\rm km\,s}^{-1}\right ) \simeq\,15{\rm km/s}$ for the asymmetric drift in the Solar Neighborhood. This is higher than the measured value of $v_a\simeq 10\,{\rm km/s}$. The difference may be due to the fact that our sample excludes stars near the midplane and therefore may overestimate $\langle v_R^2\rangle$ or due to systematic differences in other terms in the asymmetric drift formula.

We can also use our model to predict values for the Oort constants
\begin{equation}\label{eq:OortConst}
\begin{split}
    A=&\frac 12\left.\left(\frac{v_{\rm circ}}{R}-\frac{\dif v_{\rm circ}}{\dif R}\right)\right|_{R=R_0,\,z=0}\\
    B=&-\frac 12\left.\left(\frac{v_{\rm circ}}{R}+\frac{\dif v_{\rm circ}}{\dif R}\right)\right|_{R=R_0,\,z=0}~.
\end{split}
\end{equation}
The Oort constants are derived from the local value for the rotation frequency of the disk and the slope of the rotation curve. They can be used to estimate the radial contribution to the Laplacian of the gravitational potential, which in turn can be used to estimate the local matter density from a model of the vertical potential. In Paper I, we adopted $A=15.45\pm0.34\,{\rm km/s/kpc}$ and $B=-12.27\pm0.40\,{\rm km/s/kpc}$ when deriving our estimate for the local matter density. These values were obtained by averaging results from \citet{Bovy2017, Vityazev2017, Bobylev2018, Nouh2020} and \citet{Krisanova2020}. Here, we derive our own predictions for the Oort constants and present them in Table \ref{tab:Gaia_rot_curve}. We find a lower value for $A+B$ than the literature average, indicating a flatter rotation curve near the Solar Circle.
\begin{table}
    \centering
    \renewcommand{\arraystretch}{1.4}
    \begin{tabular}{lccc}
    Quantity & ES & MN & literature\\
    \hline
    $v_{\rm circ}$ & $221.6\pm0.2$ & $220.6\pm0.2$ & $219.1\pm14.1$ \\
    $\overline v_\phi$ & $211.8\pm0.2$ & $211.5\pm0.2$\\
    $A$ & $14.00\pm0.07$ & $13.81\pm0.04$ & $15.45\pm 0.34$ \\
    $B$ & $-12.69\pm0.06$ & $-12.77\pm0.04$ & $-12.27\pm 0.4$\\
    $A+B$ & $1.31\pm0.13$ & $1.04\pm0.08$ & $3.18\pm 0.52$\\
    $A-B$ & $26.69\pm0.04$ & $26.58\pm0.02$ & $27.72\pm 0.52$
    \end{tabular}
    \caption{Rotation curve quantities with their $1\sigma$ uncertainties at the Sun's position. We include the circular speed, the mean azimuthal velocity, and the Oort constants. In this table, $v_{\rm circ}$ and $\overline v_\phi$ are given in ${\rm km/s}$ while $A$, $B$, $A+B$ and $A-B$ are given in ${\rm km/s/kpc}$.}
    \label{tab:Gaia_rot_curve}
\end{table}

\subsection{Force and surface density}

\begin{figure}
	\includegraphics[width=\columnwidth]{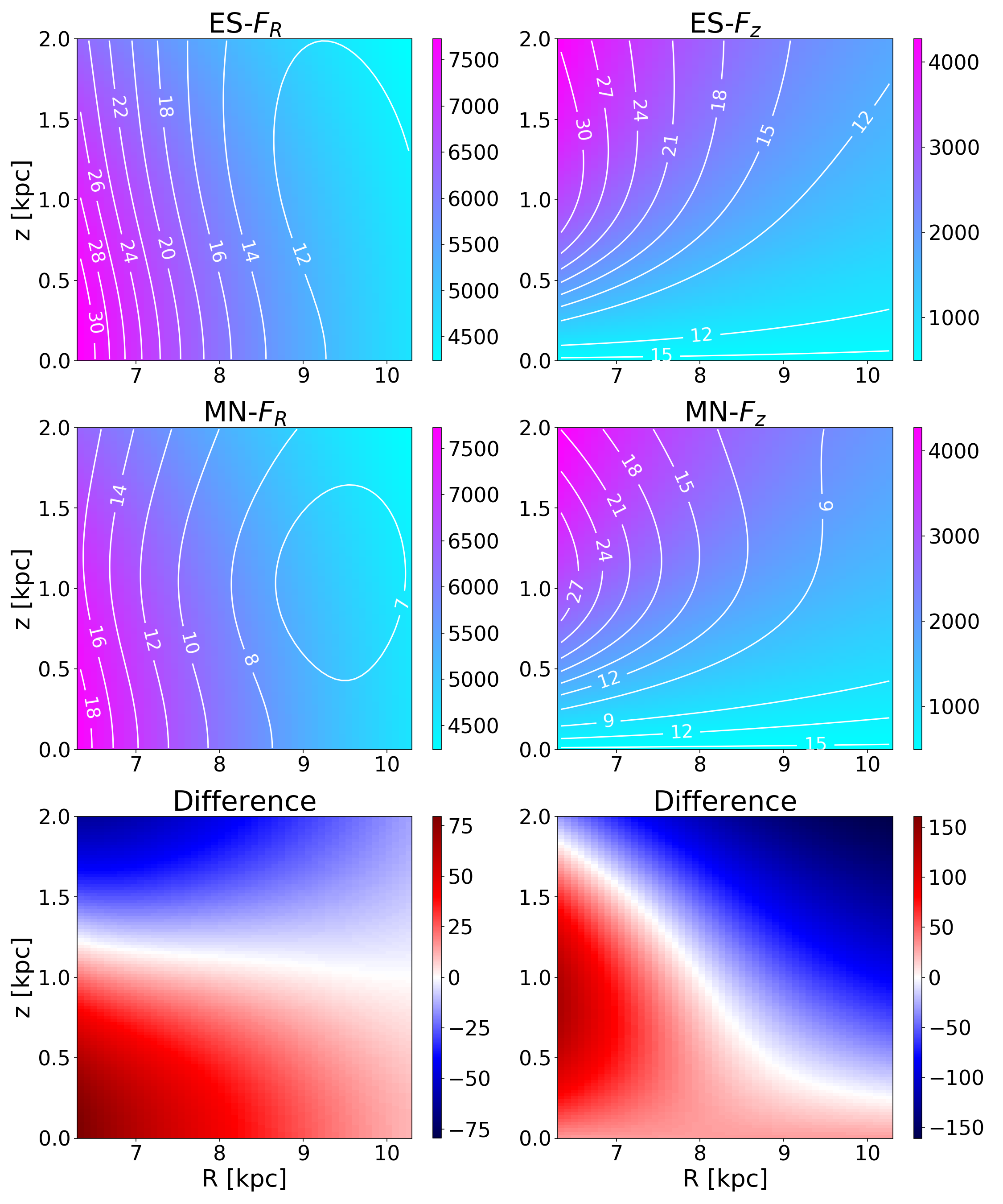}
    \caption{Model predictions for the forces from our GDR2 sample for the  ES model (top row) and MN model (middle row). Results are shown for the radial (left column) and vertical (right column) components as a color map in units of ${\rm (km/s)^2/kpc}$. We also show $1\sigma$ statistical uncertainties from the model as white contours. The bottom row shows the difference in the predictions from the ES and MN models. These panels provide an estimate of systematic errors.}
    \label{fig:Gaia_F}
\end{figure}

In Figure \ref{fig:Gaia_F}, we show model predictions of the radial and vertical components of the force in the $R-z$ plane. As was found with the mock data, the formal $1\sigma$ statistical uncertainties from the model are $\sim 5$ times smaller than the differences between the predictions from the ES and MN models. These differences again reflect the different structures of the two potentials and in particular, on the density fall-off as one moves away from the midplane. Nevertheless, the two models agree to within $\sim 1\%$ for $F_R$ and $\sim5\%$ for $F_z$ throughout the range of the sample.

\begin{figure}
	\includegraphics[width=\columnwidth]{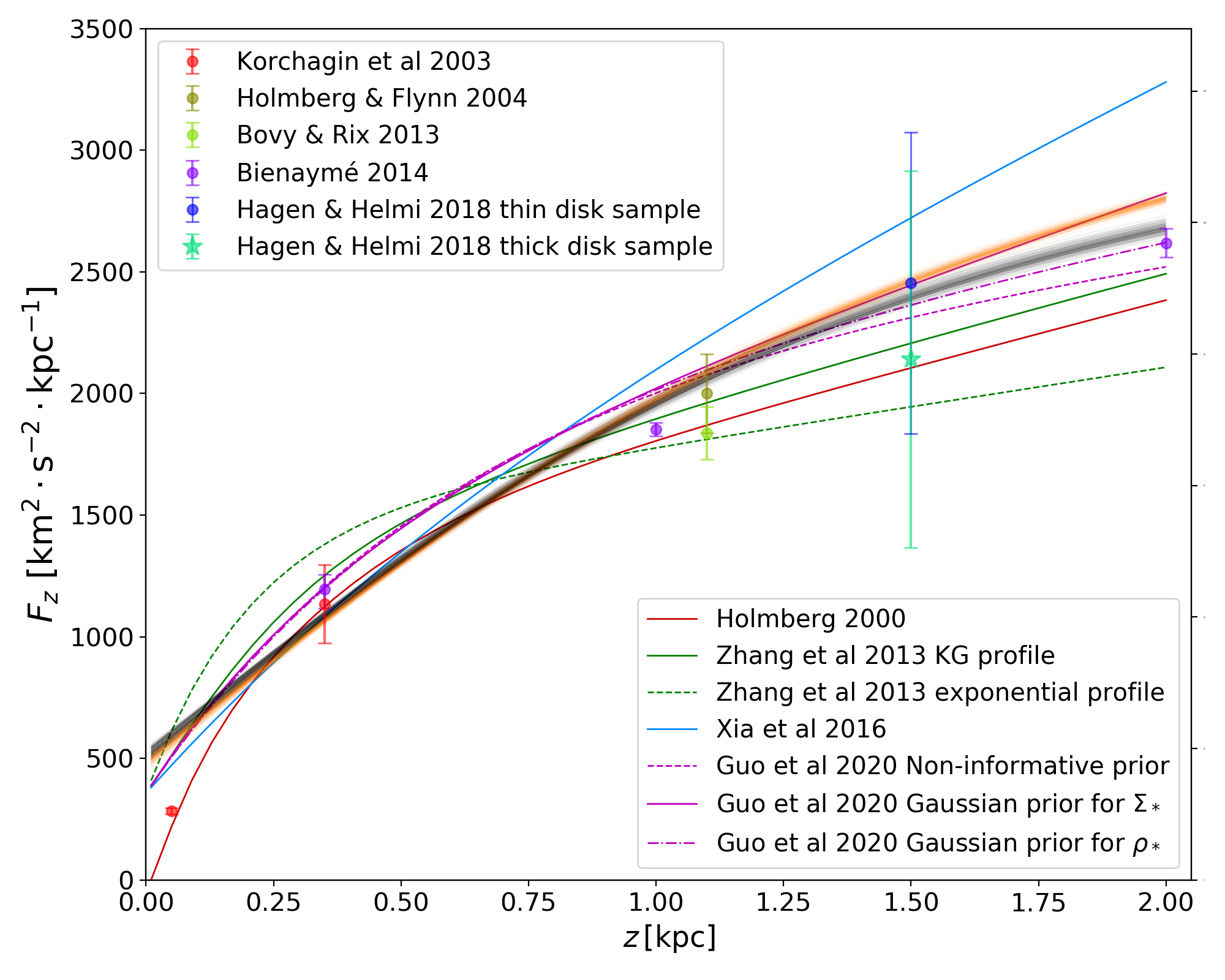}
    \caption{The vertical force in the Solar Neighborhood as predicted by our model. Also included are the same literature values as was presented in Figure 9 in Paper I. Grey curves shows predictions from 500 random samples of the MCMC chain for the ES model. Orange curves show the same for the MN model.}
    \label{fig:GaiaFz}
\end{figure}

In Figure \ref{fig:GaiaFz}, we plot the vertical force as a function of distance from the midplane at the Solar Circle. For both ES and MN models, the measured vertical force is consistent with literature values.
In the 1D approximation, the surface density is proportional to the vertical force,
\begin{equation}
    \Sigma_{\rm 1D}(z)=(2\uppi G)^{-1}F_z(z),
\end{equation}
whereas the true surface density is defined in terms of an integral over the density
\begin{equation}
    \Sigma_{\rm true}(z)\equiv\int_{-z}^z\rho(R_0,z'){\rm d}z'~.
\end{equation}
In Table \ref{tab:Fz_Sigz}, we present our predictions for both $\Sigma_{\rm 1D}$ and $\Sigma_{\rm true}$ at $z=0.5,\,1.0,\,1.5,\,2.0\,{\rm kpc}$. We see that at the Solar Circle, the deviation of $\Sigma_{\rm 1D}$ from $\Sigma_{\rm true}$ is $\lesssim2\%$.
\begin{table}
    \centering
    \renewcommand{\arraystretch}{1.4}
    \setlength{\tabcolsep}{4pt}
    \begin{tabular}{c|cc|cc}
     & \multicolumn{2}{c}{$\Sigma_{\rm 1D}\,[{\rm 10^7\,M_\odot/kpc^2}]$} & \multicolumn{2}{c}{$\Sigma_{\rm true}\,[{\rm 10^7\,M_\odot/kpc^2}]$} \\
    $z\,[{\rm kpc}]$ & ES & MN & ES & MN\\
    \hline
    $0.5$ & $4.88\pm0.04$ & $4.82\pm0.03$ & $4.76\pm0.04$ & $4.73\pm0.03$\\
    $1.0$ & $7.25_{-0.05}^{+0.06}$ & $7.31_{-0.04}^{+0.05}$ & $7.06_{-0.06}^{+0.05}$ & $7.16_{-0.05}^{+0.04}$\\
    $1.5$ & $8.87_{-0.06}^{+0.07}$ & $9.12\pm0.05$ & $8.70_{-0.07}^{+0.06}$ & $8.99\pm0.05$\\
    $2.0$ & $9.92_{-0.06}^{+0.08}$ & $10.37\pm0.04$ & $9.87_{-0.07}^{+0.06}$ & $10.35_{-0.04}^{+0.05}$
    \end{tabular}
    \caption{1D-approximated surface densities and the true surface densities at the Solar Circle for $z=0.5,1.0,1.5,2.0\,{\rm kpc}$ with their $1\sigma$ uncertainties.}
    \label{tab:Fz_Sigz}
\end{table}

\subsection{Moments of the distribution function}\label{sect:Gaia_Moment}

\begin{figure}
	\includegraphics[width=\columnwidth]{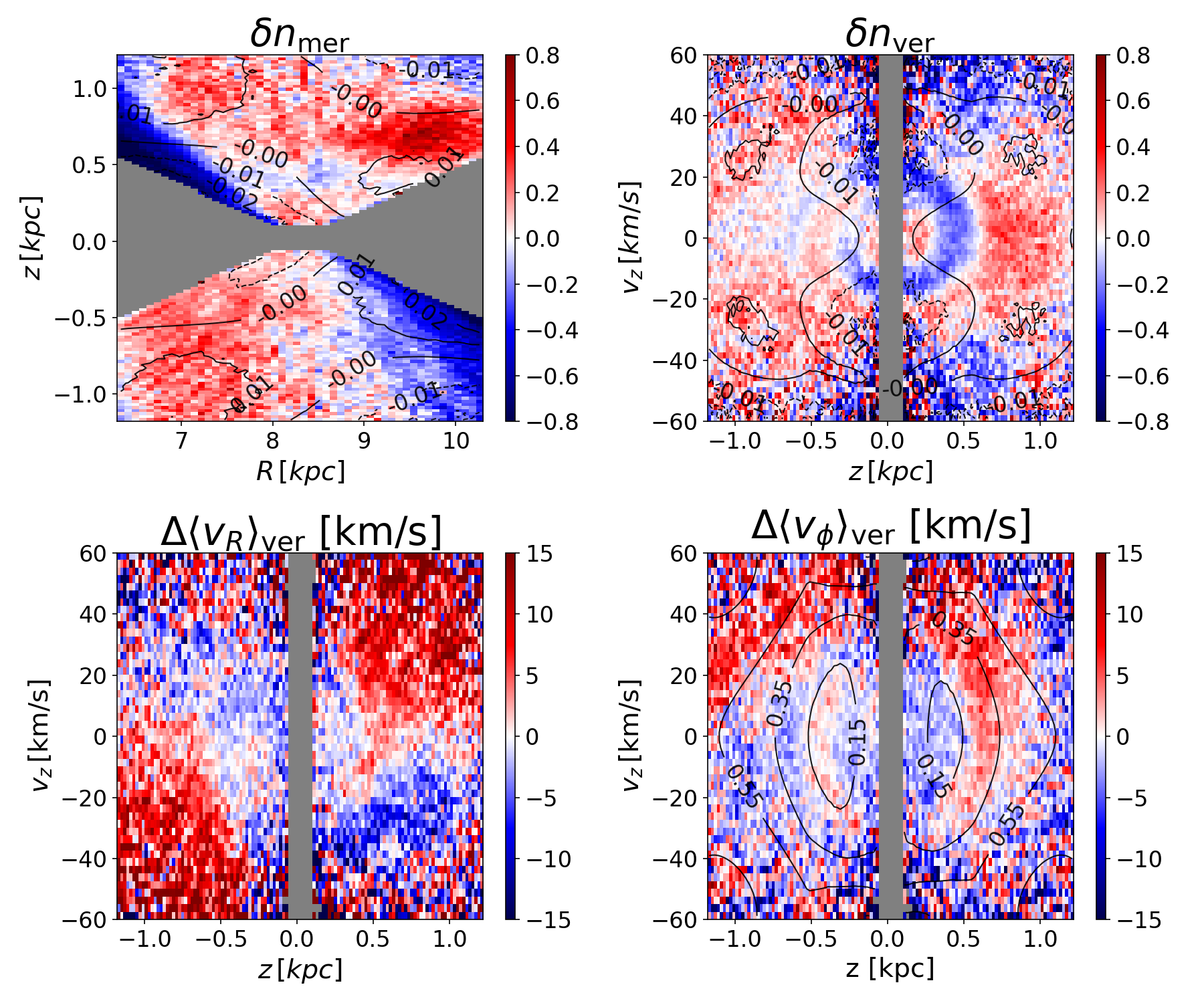}
    \caption{Residuals of the number counts and velocity moments. We show fractional residuals of number counts in the $R-z$ plane (upper left) and the $z-v_z$ plane (upper right). 
    Lower panels show data $-$ model residuals of $\langle v_R\rangle_{\rm ver}$ (left) and $\langle v_\phi\rangle_{\rm ver}$ (right). We show results for the ES model. Black contours indicate differences between predicted values for the two models. Bin sizes are the same in Figures \ref{fig:Mock_Dens} and  \ref{fig:Mock_AntojaSpiral}. We assign grey color to all bins where the ten bootstrapped samples combine to give fewer than 10 stars.}
    \label{fig:Moment_Gaia}
\end{figure}

In the upper left panel of Figure \ref{fig:Moment_Gaia}, we plot the fraction residual in number counts,
\begin{equation}
    \delta n=\frac{n_d}{n_m}-1~,
\end{equation}
in the meridonal plane. The results are shown for the ES model with contours indicating the (practically negligible) difference between ES and MN models. The relatively large residuals reflect significant departures from a plane symmetric model. This figure can be compared with results from the analysis of SDSS data by \citet{Juric2008} who also found order $\sim50\%$ departures from an equilibrium model. In particular, the prominent overdensity around $R\simeq9.5\,{\rm kpc}$ and $z\simeq0.7\,{\rm kpc}$ can be seen in their Figure 26. The pattern of over and under density predictions that run along the lower edge of the sample region may indicate that the disk is bent to negative $z$ inside the Solar Circle and positive $z$ outside the Solar Circle.

Figure \ref{fig:Moment_Gaia} also shows $\delta n$ as well as data $-$ model residuals for $\langle v_R\rangle_{\rm ver}$  and $\langle v_\phi\rangle_{\rm ver}$ in the $z-v_z$ plane. Recall that by symmetry, the model predictions for $\langle v_R\rangle$ are zero so in that panel, we are showing $\langle v_R\rangle$ from the data. The number counts panel shows the {\it Gaia} phase spiral that was discovered by \citet{antoja2018}. In the $\Delta\langle v_R\rangle_{\rm ver}$ panel, the most striking feature is a variation in bulk radial motion as one circles the phase space at a vertical energy of $300\,{\rm (km/s)^2}\lesssim E_z\lesssim1300\,{\rm (km/s)^2}$. Since the model is even in $v_R,\,v_z,$ and $z$ and our geometrical selection is independent of velocity, the variation must be intrinsic to the data. The pattern is such that the bulk motion towards the midplane is correlated with motion radially outward. At smaller $E_z$, we find hints of the phase  spirals seen in \citet{antoja2018}. We also find hints of the spirals in the $\Delta\langle v_\phi\rangle_{\rm ver}$ panel.

The spirals in the three vertical phase space panels of Figure \ref{fig:Moment_Gaia} are not as distinct or sharply defined as in \citet{antoja2018} or subsequent studies such as \citet{Li2020, Hunt2022} and our own Paper I. This is perhaps not surprising since we are considering a broad range in $R$. \citet{laporte2019,blandhawthorn2019,Li2020,Li2021,Hunt2022} and others have shown that the spiral patterns appear sharper if one bins stars according to $R$, $L_z$, Galactic azimuth angle $\phi$, etc. and that the shape of the spiral differs from one region of the Galaxy to another. Whether this reflects a change in the characteristics of the disk (e.g., surface density) or the possibility that there have been multiple disturbances in the disk, as suggested by \citet{Hunt2022} remains an open question.

\section{Residuals in frequency-angle space}\label{AAV}

It is widely accepted that the {\it Gaia} phase spirals are generated by incomplete phase mixing of a perturbation to the disk. Phase mixing is easy to understand in 1D if we ignore the back-reaction of the gravitational field generated by the perturbation on the perturbation itself, a.k.a. the self-gravity. Consider a perturbation that shifts the center of an equilibrium $z-v_z$ distribution by an amount $\Delta v_z$ so that all stars are given a velocity ``kick". Since the vertical potential is generally anharmonic such that the vertical frequency $\Omega_z$ decreases with $|z|$, the DF will be sheared into a trailing spiral. 

The process is particularly simple if we plot the evolution of the DF in $\Omega_z-\theta_z$ coordinates where $\theta_z$ is the angle associated with the vertical action. The perturbation considered above leads to a ridge of stars centered along a particular value of $\theta_z$, say $\theta_0$. In our example, the ridge runs from the origin in the $z-v_z$ plane along the positive $v_z$ axis. Phase mixing then shears the ridge into a diagonal stripe with $\theta(t) =\theta_0 + t\cdot\Omega_z$ in the $\Omega_z-\theta_z$ plane. In Paper I, and we used this idea to estimate a perturbation age of $t = 543\,{\rm Myr}$.

In what follows, we consider the model residuals in angle-frequency-angle coordinates for the full 6D phase space. The novel feature of our analysis is that the potential used to accomplish the transformation and the distribution function for the giant stars are fit simultaneously from a single data set.

\subsection{angle-action-frequency variables}\label{AAV_intro}

Actions and angles are conjugate variables of the Hamiltonian, where actions are exactly conserved in a time-independent system, which isn't true for the corresponding quantities $E_p$ and $E_z$. Furthermore, actions are adiabatic invariants. That is, they are constants of motion in time-dependent systems so long as the timescale for the potential to change is long compared to the oscillation periods. For a general axisymmetric potential,
the azimuthal action $J_\phi=L_z$ is readily explicit functions of the phase space coordinates and hence easy to determine. However, the other action-frequency-angle coordinates are implicit functions of the phase space coordinates and difficult to calculate.

In what follows, we determine action-angle variables and the associated frequencies using \texttt{AGAMA}, which employs the so-called St\"{a}kel fudge (See \citet{sanders2015} and references therein). Consider a stellar system with an axisymmetric potential $\Psi(R,z)$. A St\"{a}ckel potential is one in which there exist functions $U(u)$ and $V(v)$ and a parameter $\Delta$ such that
\begin{equation}
    \Psi(R,z)=\frac{U(u)-V(v)}{\sinh^2 u+\sin^2 v}
\end{equation}
with
\begin{equation}
    R=\Delta\sinh u\sin v,\qquad z=\Delta\cosh u\cos v
\end{equation}
In a St\"{a}ckel potential, stellar orbits are separable in $u$ and $v$ and all three pairs of angle-action variables and the associated frequencies $(J_i,\,\theta_i,\Omega_i)$ ($i=u,\,v,\,\phi$) can be calculated exactly, where frequencies are the time derivatives of the associated angles which conserves for any individual star. The idea of the St\"{a}ckel fudge is to extend this result to general potentials by finding an approximation of the St\"{a}ckel form for the gravitational potential over small region of the Galaxy. In this way, subscripts $u$ and $v$ will be replaced with $R$ and $z$.

In principle, a perturbation to the disk will lead to a pattern of diagonal stripes in the $\Omega_i-\theta_i$ planes provided the potential is constant after the initial kick and self-gravity is negligible. As we will see, the pattern of residuals is far more complex.

\subsection{residuals in action-frequency-angle coordinates}\label{AAV_res}

It is straightforward to calculate action-frequency-angle coordinates from ${\boldsymbol r}$ and ${\boldsymbol v}$ using our best-fit potential since the St\"{a}ckel Fudge has been implemented for the potentials used in our model within the \texttt{AGAMA} toolbox \citep{agama}. It is more difficult to calculate model predictions for the number counts in terms of action-frequency-angle variables since this requires computationally expensive calculation of the Jacobians for the transformations. In what follows, we use a Monte Carlo approach. We first generate 50 million particles uniformly across the 6D phase space with the constraints $|\phi|<4^\circ$, $|R-R_0|<2\,{\rm kpc}$ and $|z-z_0|<2\,{\rm kpc}$. Each particle is assigned a weight proportional to $f(\boldsymbol r,\boldsymbol v)g(\boldsymbol r)$ where the normalization constant is determined such that the sum of weights for all stars equals the average number of stars in our 10 bootstrapped data sets. We then convert the phase space coordinates to frequencies and angles using \texttt{AGAMA}. The model-predicted distribution of stars in the space of any two quantities is just the weighted two-dimensional histogram of these particles. Similarly, we combine our 10 bootstrapped {\it Gaia} data sets together and assign each star a weight of $0.1$ to arrive at the same quantities for data.

\begin{figure}
  \centering
  \includegraphics[width=\columnwidth]{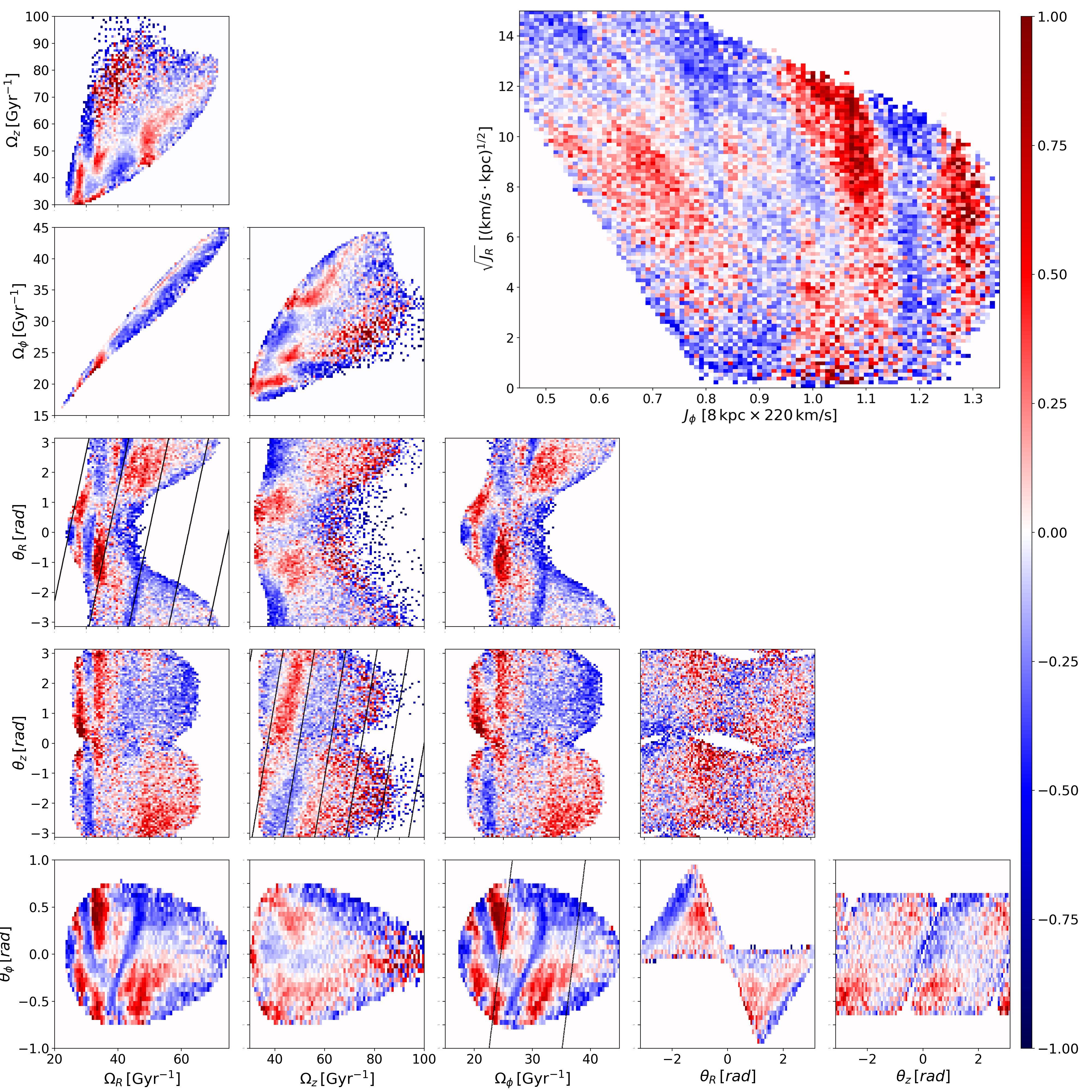}
  \caption{Projections of the fractional residuals in star counts in the spaces of all 15 coordinate pairs from the six frequency-angle coordinates for the ES model. Note that $\theta_\phi$ does not span the entire $(-\uppi,\,\uppi)$ range due to our geometrical selection. We plot straight lines corresponding to $t=500\,{\rm Myr}$ in the $\Omega_R-\theta_R$, $\Omega_z-\theta_z$, and $\Omega_\phi-\theta_\phi$ panels. Also shown are the fractional residuals in the $J_\phi-\sqrt{J_R}$ plane. The scaling for this panel is chosen to match Figure 5 of \citet{Trick2019}. Only pixels with at least 5 stars are shown.}
  \label{fig:Gaia_AAV}
\end{figure}

\begin{figure}
  \centering
  \includegraphics[width=\columnwidth]{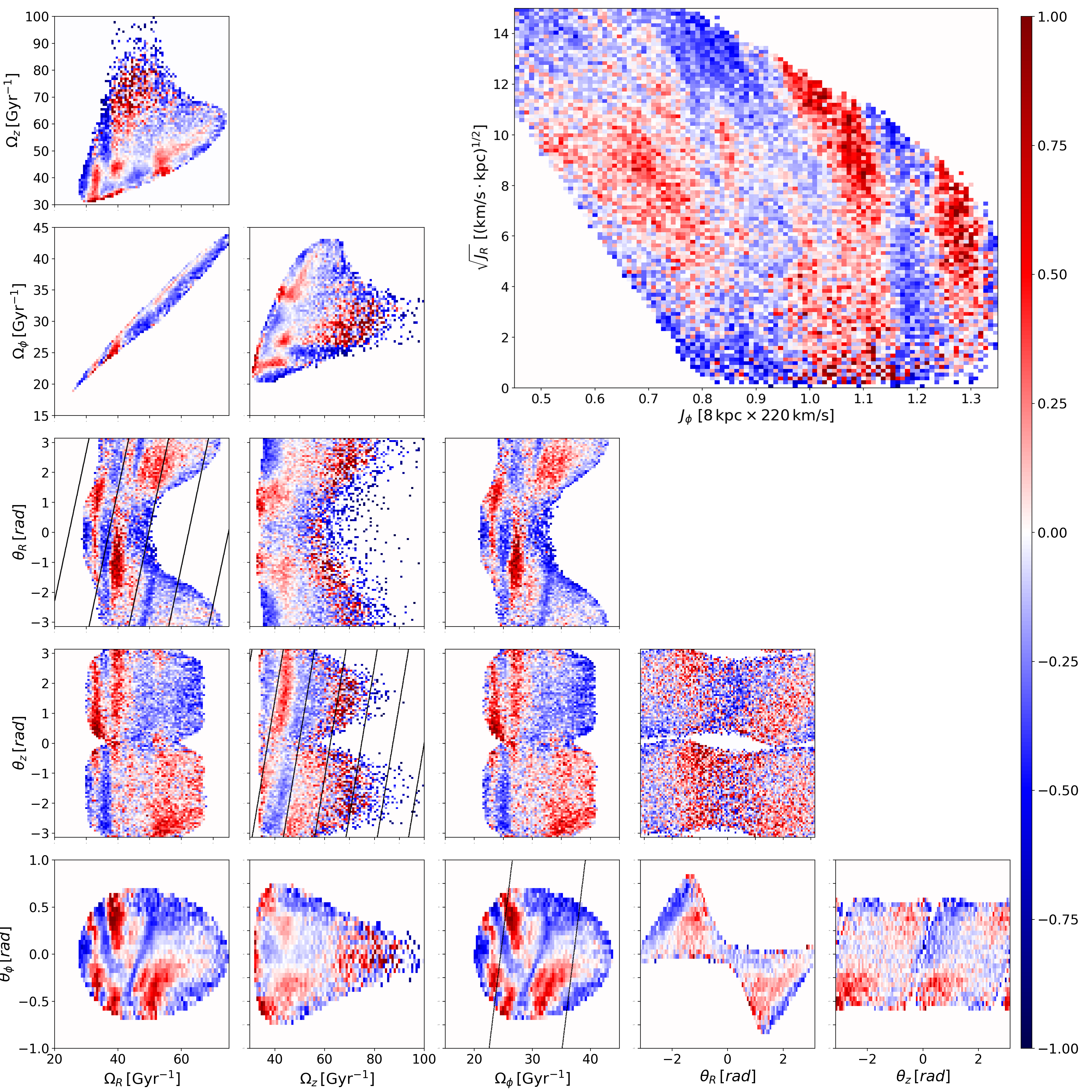}
  \caption{Same as Figure \ref{fig:Gaia_AAV} but for the MN model.}
  \label{fig:Gaia_AAV_MN}
\end{figure}

In Figure \ref{fig:Gaia_AAV}, we show the fractional residuals for the number counts in the spaces of all 15 pairs of the six frequency-angle coordinates, as well as in the $J_\phi-\sqrt{J_R}$ space, for the ES model. We show the same for the MN model in Figure \ref{fig:Gaia_AAV_MN}. The span of frequencies are a little different but the residual maps are very similar and reveal a wealth of substructures. The diagonal stripes in the $\Omega_z-\theta_z$ space are analogous to those seen in the lower panel of Figure 12 in Paper I. The slope of the stripes gets shallower with increasing $\Omega_z$. For the MN model, the slope for $\Omega_z\lesssim 45\,{\rm Gyr}^{-1}$ is consistent with a perturbation age of $500\,{\rm Myr}$, which is similar to what we found in Paper I. For the ES model though, the slope is clearly smaller. For larger $\Omega_z$, the slope for both models are consistent with an age of $\sim 250\,{\rm Myr}$, though it is difficult to make precise interpretations since the stripes are somewhat disjoint. The structures in the $\Omega_R-\theta_R$ are even more challenging to interpret. In particular, it is difficult to find diagonal features that extend across of a wide range in $\theta_R$ though there are certainly features consistent with a disturbance that occurred $500\,{\rm Myr}$ ago for both models. The implication is that the perturbation (or multiple perturbations) to the disk has a strong dependence on radial action.
Finally, we come to the $\Omega_\phi-\theta_\phi$ plane. Here, the difficulty is that we have results across a limited range in $\phi$. In principle, residuals in this plane should encode some of the variations in the phase spirals as a function of $J_z$ and $\theta_z$ as see in \citet{Hunt2022} and others.

We next turn to the $\Omega_R-\Omega_\phi$ plane. We see that most of the stars in our sample lie along a narrow ridge in this plane, given that we don't show pixels with fewer than 5 stars. The ridge is roughly defined by the condition $\frac 43\lesssim\frac{\Omega_\phi}{\Omega_R}\lesssim\frac 53$, which is as expected since $\frac{\Omega}{\kappa}\simeq\sqrt 2$ for flat rotation curve and small epicyclic motions.

The more useful phase space projection is the $J_\phi-\sqrt{J_R}$ plane, which is shown in the upper right of Figure \ref{fig:Gaia_AAV}. The distribution of stars in this plane has been studied in the context of moving groups by \citet{Trick2019} and the scaling in our figure is chosen to match their Figure 5. As those authors note, the appearance of moving groups in their figure is strongly affected by selection effects (see their Figure 2).
The over-density of particles in Figure \ref{fig:Gaia_AAV} at $(J_\phi,\,\sqrt{J_R}) = (1,2)$ in scaled units is likely a combination of the Hyades and Coma Berenices moving groups whereas the structure extending to higher $\sqrt{J_R}$ and lower $J_\phi$ may be the Hercules moving group. Finally, there are the near-vertical stripes at higher $\sqrt{J_R}$, which are likely connected to the velocity arches discovered \citet{gaiadr2_vrad} as discussed in \citet{Trick2019} though the extension of the most prominent over-density to lower $J_R$ may also include stars from the Sirius moving group.


\section{Discussion}\label{Discussion}

One troubling aspect of our analysis is that the statistical uncertainties appear to under-represent the true uncertainties of the model. In our mock data tests, we found this to be about an order of magnitude smaller than the true error. The implication is that the model is over-fitting the data, which is perhaps not surprising given that we have measurements for six phase space coordinates for $\sim260$ thousand stars and an 18-parameter model. More to the point, the functional form of the potential is likely too restrictive. This conjecture is supported by the fact that the errors found in our mock data test are comparable to the differences between results from two different choices for the potential. For the purposes of this paper, we therefore advocate using this difference to estimate systematic uncertainties. We stress that these systematic uncertainties are still impressively small. Moving forward, we will consider more flexible forms for the potential, either by adding in additional components or using an expansion in a set of basis functions as in the self-consistent field method.

Our choice for the DF also deserves further consideration. In this paper, we assumed that the stars in the sample could be decomposed into thin and thick components. Alternatively, in Paper I, we introduced the rational linear distribution function (RLDF), which described a superposition of components with smoothly varying scale height and velocity dispersion. This model correctly predicted the differential surface density profile, that is, surface density in the Solar Neighborhood as a function of vertical velocity dispersion (see Figure 11 of Paper I and Figure 8 of \citet{bovy2012}). At present, we do not have a three-integral extension of the RLDF but it may be worthwhile to explore whether one exists.

In Paper I, we computed the likelihood function by first binning the data in the $z-v_z$ plane and then computing a $\chi^2$-statistic based on number counts in these bins. Here, we compute the likelihood function directly from the unbinned data by taking the product of the DF at the measured phase space positions of each of the stars in our sample. Binning in the full 6D phase space is clearly unfeasible with $\sim260$ thousand stars. Even with two orders of magnitude more stars, which will be the case after the Third and Fourth {\it Gaia} data releases, the number of stars will be too small to provide statistically meaningful averages on a 6D grid. On the other hand, the increase in the number of stars may put the computational cost of our present method out of reach. An intermediate approach is to bin stars in the $R-z-v_z$ sub-space of the entire phase space and base the likelihood function on number counts as well as low-order $v_R$ and $v_\phi$ moments of the DF.

Perhaps the most interesting result from this work is the complexity of structure seen when residuals in the number count are plotted in terms of frequency-angle or action-angle variables. Some of the features in these plots can be identified with known moving groups or the velocity arches discovered by \citet{gaiadr2_vrad} as in \citet{Trick2019}. We also identify the phase spirals discovered by \citet{antoja2018}. The complexity of these structures suggests that the Solar Neighborhood may have experienced multiple disturbances and the associated phase space perturbations can depend on both the action (or frequency) and angle of individual stars. Moreover, self-gravity can also influence the evolution of perturbations as was shown in the case of the phase spirals by \cite{Darling2019}. Numerical simulations combined with techniques such as the St\"{a}ckel fudge will provide an invaluable tool for understanding the complicated dynamics of the Solar Neighborhood \citep{Laporte2018, laporte2019, Li2021, bennett2022, GarciaConde2022, Hunt2022}. There is also the question of the interplay between disk perturbations and our attempts to determine the local gravitational potential and dark matter density \citep{banik2017, Widmark1, Widmark2, Widmark3, Sivertsson2022}.

\section{Conclusion}\label{Conclusion}

In this work, we simultaneously determine the DF for a tracer population and the Milky Way gravitational potential within $\sim 2\,{\rm kpc}$ of the Sun using astrometric data of a sample of giant stars from GDR2. The results are used to predict the radial and vertical components of the force in the sample region. We consider two different models for the contribution of the disk to the gravitational potential and use the difference in predictions from the two models as a means of estimating systematic errors. Since the model implicitly includes the effect of asymmetric drift, we are able to make separate predictions for the mean azimuthal velocity curve and the circular speed curve. Our value for the circular speed in the Solar Neighborhood is consistent with literature average though the rotation curve is somewhat flatter than the curves usually found in the literature. Our measured vertical forces are consistent with literature values, while the radial and vertical forces for our entire sample range agrees to $\sim1\%$ and $\sim 5\%$ respectively between two models.

An attractive feature of our method is that the residuals of the DF can be viewed in either the original phase space coordinates or in terms of angle-frequency-angle coordinates. In original phase space coordinates, residuals in number counts in the $R-z$ plane reveal complicated substructures that indicate the departure of the Solar Neighborhood from dynamical equilibrium. The number counts residuals in the $z-v_z$ plane show spiral-like patterns similar to what was found by \citet{antoja2018} and also seen in Paper I, though the patterns here are more disorganized and faint. The difference may be due to the fact that we are considering a wider range in Galactocentric radius or that the potential differs from the true one. Similar results are obtained for the mean radial and vertical velocity components in the $z-v_z$ plane, which also shows coupling between in-plane and vertical motions of stars. When plotted in angle, frequency, and angle coordinates, the star count residuals reveal a wide range of complicated structures. These include diagonal stripes that suggest phase mixing of disturbances $250-500\,{\rm Myr}$ ago, moving groups, and velocity arches.

\section*{Data availability}

The {\it Gaia} Second Data Release is available at the following website: https://gea.esac.esa.int/archive/. All other data used for our work is available through the links posted in the footnotes where necessary.

\section*{Acknowledgements}

We acknowledge the financial support of the Natural Sciences and Engineering Research Council of Canada.

\appendix

\section{Normalization factor for DF}\label{app:norm}

From Equations \ref{eq:DF}, \ref{eq:DF_indv} and \ref{eq:norm_form}, one can see that
\begin{equation}
    \mathcal N=\eta\mathcal N_1+(1-\eta)\mathcal N_2
\end{equation}
where ($i=1,2$)
\begin{equation}
    \mathcal N_i\equiv\int\frac\Omega\kappa\frac{\rhot g({\boldsymbol r})}{\sigmat_{Ri}^2\sigmat_{zi}}\exp{\left(-\frac{E_p-E_c}{\sigmat_{Ri}^2}-\frac{E_z}{\sigmat_{zi}^2}\right)}\dif^3{\boldsymbol r}\dif^3{\boldsymbol v}
\end{equation}
We attempt to calculate this integral in Galactocentric cylindrical coordinates. First of all, the only terms that involves $v_z$ and $v_R$ respectively are $E_z$ as Equations \ref{eq:Ez} and $E_p-E_c$ which follows:
\begin{equation}
    E_p-E_c=\frac 12v_R^2+\Psieff(R,0)-\Psieff(R_c,0)\geq0
\end{equation}
where
\begin{equation}\label{eq:Psieff}
    \Psieff(R,z)\equiv\Psi(R,z)+\frac{{L_z}^2}{2R^2}
\end{equation}
is the effective potential. Integrating these dimensions out, we have
\begin{equation}
\begin{split}
    \mathcal N_i=2\uppi&\int\frac\Omega\kappa\frac{\rhot}{\sigmat_{Ri}}g({\boldsymbol r})\times\\
    &\exp{\left[-\frac{\DPsieff(R,R_c)}{\sigmat_{Ri}^2}-\frac{\DPsiz(R,z)}{\sigmat_{zi}^2}\right]}\dif^3{\boldsymbol r}\dif v_\phi
\end{split}
\end{equation}
where
\begin{equation}
\begin{split}
    &\DPsieff(R,R_c)\equiv\Psieff(R,0)-\Psieff(R_c,0)\\
    &\DPsiz(R,z)\equiv\Psi(R,z)-\Psi(R,0)
\end{split}
\end{equation}

Then, one should observe that $\dif^3{\boldsymbol r}=R\dif R\dif\phi\dif z$ and that $g({\boldsymbol r})$ is the only term that involves $\phi$. Integrating the $\phi$-dimension out, we have:
\begin{equation}\label{eq:I_inter}
\begin{split}
    \mathcal N_i=2\cdot2\uppi&\int_0^{+\infty}\dif R\int_{-\infty}^{+\infty}\dif z\tilde g(R,z)\int_0^{+\infty}\frac\Omega\kappa\frac{\rhot}{\sigmat_{Ri}}\times\\
    &\exp{\left[-\frac{\DPsieff(R,R_c)}{\sigmat_{Ri}^2}-\frac{\DPsiz}{\sigmat_{zi}^2}\right]}(R\dif v_\phi)
\end{split}
\end{equation}
where 
\begin{equation}
    \tilde g(R,z)\equiv\int_0^{2\pi} g(\boldsymbol r)\dif\phi
\end{equation}
indicates the effect of geometrical selection in the $R-z$ plane, and the additional factor of two comes from the symmetry of the integrand with regard to $v_\phi$. For the geometrical selection applied in our work as described in Section \ref{DataSel},
\begin{equation}
    \tilde g(R,z)=2\min\left\{\phi_m,\,\arccos\frac
    {{R_0}^2+R^2-\left(\frac{z-z_0}{\tan 15^\circ}\right)^2}{2R_0R}\right\}
\end{equation}
if $80\,{\rm pc}<|z-z_0|<2\,{\rm kpc}$ and
\begin{equation*}
    |R-R_0|<\min\left\{2\,{\rm kpc},\,\frac{|z-z_0|}{\tan 15^\circ}\right\}
\end{equation*}
Otherwise, $\tilde g(R,z)=0$.

To evaluate $\mathcal N_i$ from Equation \ref{eq:I_inter}, one needs to write $\dif v_\phi$ in terms of $R_c$ assuming constant position coordinates. To do this, we first recognize that the epicyclic frequencies as functions of $R_c$ are calculated as
\begin{equation}\label{eq:Omega}
    \Omega(R_c)=\frac{v_c}{R_c}=\frac{L_z}{R_c^2}
\end{equation}
and
\begin{equation}\label{eq:kappa}
    \kappa^2(R_c)=\frac{\partial^2 \Psieff(R_c,0)}{\partial R_c^2}
    =\frac{\partial^2\Psi(R_c,0)}{\partial R_c^2}+\frac{3L_z^2}{R_c^4}
\end{equation}
We define the following two derivatives:
\begin{equation}
    D_k\equiv\left.\frac{\partial^k \Psi}{\partial R^k}\right|_{R=R_c,\,z=0},\qquad k=1,2
\end{equation}
For $k=1$, given the effective potential as in Equation \ref{eq:Psieff} and that $\left.\frac{\partial \Psieff}{\partial R}\right|_{R=R_c,\,z=0}=0$, we have:
\begin{equation}
    D_1=-\left.\frac{\dif}{\dif R}\left(\frac{{L_z}^2}{2R^2}\right)\right|_{R=R_c}=\frac{{L_z}^2}{{R_c}^3}=R_c\Omega^2
\end{equation}
For $k=2$, Equation \ref{eq:kappa} indicates that
\begin{equation}
    D_2=\kappa^2-3\frac{{L_z}^2}{{R_c}^4}=\kappa^2-3\Omega^2
\end{equation}
Note that
\begin{equation}
    Rv_\phi=L_z=R_cv_c={R_c}^{\frac 32}{D_1}^{\frac 12}
\end{equation}
Therefore,
\begin{equation}\label{eq:dvphi_dRc}
\begin{split}
    R\left(\frac{\partial v_\phi}{\partial R_c}\right)_{\vec{r}}
    =&\left(\frac 32{R_c}^{\frac 12}\right){D_1}^{\frac 12}+
           {R_c}^{\frac 32}\left(\frac 12{D_1}^{-\frac 12}D_2\right)\\
    =&\frac 32{R_c}^{\frac 12}\left(R_c\Omega^2\right)^{\frac 12}+
    \frac {{R_c}^{\frac 32}\left(\kappa^2-3\Omega^2\right)}
    {2\left(R_c\Omega^2\right)^{\frac 12}}\\
    =&\frac{R_c\kappa^2}{2\Omega}
\end{split}
\end{equation}
Plugging this into Equation \ref{eq:I_inter}:
\begin{equation}
\begin{split}
    \mathcal N_i=2\uppi&\int_0^{+\infty}\dif R\int_{-\infty}^{+\infty}\dif z\tilde g(R,z)
    \int_0^{+\infty}\frac{R_c\kappa\rhot_i}{\sigmat_{Ri}}\times\\
    &\exp{\left[-\frac{\DPsieff(R,R_c)}{\sigmat_{Ri}^2}-\frac{\DPsiz(R,z)}{\sigmat_{zi}^2}\right]}\dif R_c
\end{split}
\end{equation}

\section{Moments of the DF}\label{app:dens_prof}

For density moments, one should observe from Equations \ref{eq:DF}, \ref{eq:nRz} and \ref{eq:nperp} that
\begin{subequations}
\begin{align}
    n_{\rm mer}(R,z)=&\mathcal N_{\rm mer}\left[\eta \tilde n_{\rm mer,1}+(1-\eta)\tilde n_{\rm mer,2}\right]\\
    n_{\rm ver}(z,v_z)=&\mathcal N_{\rm ver}\left[\eta \tilde n_{\rm ver,1}+(1-\eta)\tilde n_{\rm ver,2}\right]
\end{align}
\end{subequations}
where $\mathcal N_{\rm mer}$ and $\mathcal N_{\rm ver}$ are normalization factors mentioned in Section \ref{moment_sect}. For each individual disk ($i=1,2$):
\begin{subequations}
\begin{align}
    \tilde n_{{\rm mer},i}=\tilde n_{{\rm mer},i}(R,z)=&\int f_i({\boldsymbol r},\,{\boldsymbol v})g({\boldsymbol r})(R\dif\phi)\dif^3{\boldsymbol v}\\
    \tilde n_{{\rm ver},i}=\tilde n_{{\rm ver},i}(z,v_z)=&\int f_i({\boldsymbol r},\,{\boldsymbol v})g({\boldsymbol r})(R\dif R\dif\phi)\dif v_R\dif v_\phi
\end{align}
\end{subequations}
As for $v_\phi$ moments, one should observe from Equations \ref{eq:vphimean_Rz} and \ref{eq:vphiperp} that
\begin{subequations}
\begin{align}
    \langle v_\phi\rangle_{\rm mer}(R,z)=&\frac{\eta \xi_1+(1-\eta)\xi_2}{\eta\tilde n_{\rm mer,1}+(1-\eta)\tilde n_{\rm mer,2}}\\
    \langle v_\phi\rangle_{\rm ver}(z,v_z)=&\frac{\eta \zeta_1+(1-\eta)\zeta_2}{\eta \tilde n_{\rm ver,1}+(1-\eta)\tilde n_{\rm ver,2}}
\end{align}
\end{subequations}
where
\begin{subequations}
\begin{align}
    \xi_i=\xi_i(R,z)=&\int v_\phi f_i({\boldsymbol r},{\boldsymbol v})g({\boldsymbol r})(R\dif\phi)\dif^3{\boldsymbol v}\\
    \zeta_i=\zeta_i(z,v_z)=&\int v_\phi f_i({\boldsymbol r},{\boldsymbol v})g({\boldsymbol r})(R\dif R\dif\phi)\dif v_R\dif v_\phi
\end{align}
\end{subequations}

To get the profiles we need, we need to evaluate integrals $\tilde n_{{\rm mer},i}(R,z)$, $\tilde n_{{\rm ver},i}(z,v_z)$, $\xi_i(R,z)$ and $\zeta_i(z,v_z)$. For all these integrals, one should integrate $v_R$ and $\phi$ out in the same way as in Appendix \ref{app:norm}, and plug in $v_\phi=\frac{L_z}{R}$ for $v_\phi$ and Equation \ref{eq:dvphi_dRc} for $R\dif v_\phi$. Then, for $\tilde n_{{\rm mer},i}(R,z)$ and $\xi_i(R,z)$ we should further integrate out $v_z$ also in the same way as in Appendix \ref{app:norm}. When the smoke clears:
\begin{equation}
\begin{split}
  \tilde n_{{\rm mer},i}(R,z)=&2\uppi\tilde g(R,z)\int_0^{+\infty}\dif R_c \frac{R_c\kappa\rhot_i}{\sigmat_{Ri}}\times\\
    & \exp{\left[-\frac{\DPsieff(R,R_c)}{\sigmat_{Ri}^2}-\frac{\DPsiz(R,z)}{\sigmat_{zi}^2}\right]}
\end{split}
\end{equation}
\begin{equation}
\begin{split}
  \tilde n_{{\rm ver},i}(z,v_z)=&\sqrt{2\uppi}\int_0^{+\infty}\dif R_c \frac{R_c\kappa\rhot_i}{\sigmat_{Ri}\sigmat_{zi}}\int_0^{+\infty}\dif R\tilde g(R,z)\times\\
    &\exp{\left[-\frac{\DPsieff(R,R_c)}{\sigmat_{Ri}^2}-\frac{E_z}{\sigmat_{zi}^2}\right]}
\end{split}
\end{equation}
\begin{equation}
\begin{split}
    \xi_i(R,z)=&\frac{2\uppi\tilde g(R,z)}R\int_0^{+\infty}\dif R_c \frac{L_zR_c\kappa\rhot_i}{\sigmat_{Ri}}\times\\
    & \exp{\left[-\frac{\DPsieff(R,R_c)}{\sigmat_{Ri}^2}-\frac{\DPsiz(R,z)}{\sigmat_{zi}^2}\right]}
\end{split}
\end{equation}
\begin{equation}
\begin{split}
  \zeta_i(z,v_z)=&\sqrt{2\uppi}\int_0^{+\infty}\dif R_c \frac{L_zR_c\kappa\rhot_i}{\sigmat_{Ri}\sigmat_{zi}}\int_0^{+\infty}\dif R \frac{\tilde g(R,z)}R\times\\
    &\exp{\left[-\frac{\DPsieff(R,R_c)}{\sigmat_{Ri}^2}-\frac{E_z}{\sigmat_{zi}^2}\right]}
\end{split}
\end{equation}
where $E_z=E_z(R,z,v_z)$ follows Equations \ref{eq:Ez}.



\bibliographystyle{mnras}
\bibliography{PaperII}


\bsp	
\label{lastpage}
\end{document}